\RequirePackage{fix-cm}
\documentclass[twocolumn]{svjour3}          
\smartqed  
 \usepackage{microtype}
\usepackage{lipsum}
\usepackage[htt]{hyphenat}
\usepackage{hyperref}

\usepackage{pgf-pie}
\usepackage{subfiles}
\usepackage{graphicx}
\usepackage[utf8]{inputenc}
\usepackage{comment}
\usepackage[english]{babel}
\usepackage{amssymb}
\usepackage{textcomp}
\usepackage{adjustbox}
\usepackage{indentfirst}
\usepackage{mathtools}
\usepackage[usestackEOL]{stackengine}
\usepackage{amsmath}
\usepackage{ltxtable}
\usepackage{algorithm}
\usepackage{algpseudocode}
\usepackage{textcomp}
\usepackage{mathpazo}
\usepackage{tgpagella}
\usepackage{color,soul}
\usepackage{dirtytalk}
\usepackage{tcolorbox}
\usepackage{graphicx,subcaption,chngpage,calc}
 \usepackage[math]{cellspace}
  \usepackage{ltablex}
\usepackage{xtab, lipsum, array}
\usepackage{caption}
\usepackage{multirow}
\usepackage{tabularx,ragged2e,booktabs,caption}
\newcolumntype{L}{>{\raggedright\arraybackslash}X}

\stackMath

\DeclareMathSizes{12}{17.28}{9}{7}
\usepackage{color}
\usepackage{multirow, makecell}
  \usepackage{ltablex}

\usepackage{url}

\usepackage{tabularx}
\usepackage{pgfplots}
\pgfplotsset{width=7cm,compat=1.8,tick label style={font=\small}}
\usepackage{pgfplotstable}
\usepackage{sfmath}
\usepackage{array, booktabs, makecell}
\usepackage{color}
\usepackage{csquotes}
\usepackage{url}
\usepackage{comment}
\usepackage{tikz}
\usepackage{balance}
\usepackage{listings}
\usepackage{breqn}
\usepackage{booktabs} 
\usepackage{soul} 
\usetikzlibrary{fit} 
\usetikzlibrary{positioning}
\usetikzlibrary{arrows}
\usetikzlibrary{shapes.multipart}
\usepackage{multirow} 
\usepackage{multicol}
\usepackage{varwidth} 

\definecolor{Gray}{gray}{0.80} 

\usepackage{mathtools}
\usepackage{amsmath}

\makeatletter
\def\BState{\State\hskip-\ALG@thistlm}
\makeatother

\usepackage{listings}

\newcommand{\lstbg}[3][0pt]{{\fboxsep#1\colorbox{#2}{\strut #3}}}
\lstdefinelanguage{diff}{
  basicstyle=\ttfamily\scriptsize,
  morecomment=[f][\lstbg{red!20}]-,
  morecomment=[f][\lstbg{green!20}]+,
  morecomment=[f][\textit]{@@},
}

\definecolor{javared}{rgb}{0.6,0,0} 
\definecolor{javagreen}{rgb}{0.25,0.5,0.35} 
\definecolor{javapurple}{rgb}{0.5,0,0.35} 
\definecolor{javadocblue}{rgb}{0.25,0.35,0.75} 
 
\usepackage{tikz,xcolor,hyperref}
\definecolor{lime}{HTML}{A6CE39}
\DeclareRobustCommand{\orcidicon}{%
	\begin{tikzpicture}
	\draw[lime, fill=lime] (0,0) 
	circle [radius=0.16] 
	node[white] {{\fontfamily{qag}\selectfont \tiny ID}};
	\draw[white, fill=white] (-0.0625,0.095) 
	circle [radius=0.007];
	\end{tikzpicture}
	\hspace{-2mm}
}

\foreach \x in {A, ..., Z}{%
	\expandafter\xdef\csname orcid\x\endcsname{\noexpand\href{https://orcid.org/\csname orcidauthor\x\endcsname}{\noexpand\orcidicon}}
}

\begin{document}

\title{Refactoring for Reuse: An Empirical Study
}


\author{Eman Abdullah AlOmar \orcidA{} \and  Tianjia Wang  \and Vaibhavi Raut \and Mohamed Wiem Mkaouer \orcidB{} \and Christian Newman \orcidF{}  \and Ali Ouni \orcidC{}  
}

\authorrunning{AlOmar et al.}

\institute{%
  Eman Abdullah AlOmar \and  Tianjia Wang  \and Vaibhavi Raut \and Mohamed Wiem Mkaouer \and Christian Newman 
  \at
  Rochester Institute of Technology \\
  \email{\{eaa6167,twang,vraut,mwmvse,cdnvse\}@rit.edu}           
  \and
  Ali Ouni
  \at
  ETS Montreal, University of Quebec \\
  \email{ali.ouni@etsmtl.ca}
}

\date{Received: date / Accepted: date}

\maketitle

\begin{abstract}
 Refactoring is the \textit{de-facto} practice to optimize software health. While several studies propose refactoring strategies to optimize software design through applying design patterns and removing design defects, little is known about how developers actually refactor their code to improve its reuse. Therefore, we extract, from 1,828 open source projects, a set of refactorings that were intended to improve the software reusability. We analyze the impact of reusability refactorings on the state-of-the-art reusability metrics, and we compare the distribution of reusability refactoring types, with the distribution of the remaining mainstream refactorings. Overall, we found that the distribution of refactoring types, applied in the context of reusability, is different from the distribution of refactoring types in mainstream development. In the refactorings performed to improve reusability, source files are subject to more design level types of refactorings. Reusability refactorings significantly impact, high-level code elements, such as packages, classes, and methods, while typical refactorings, impact all code elements, including identifiers, and parameters. \textcolor{black}{These findings provide practical insights into the current practice of refactoring in the context of code reuse involving the act of refactoring.}
\end{abstract}

\section{Introduction} \label{sec:introduction}

Refactoring is defined as the process of changing software system in such way that changes improve software quality and do not alter the software behaviour~\cite{opdyke1992refactoring,fowler2018refactoring,alomar2021preserving}. 
Refactoring is one of the commonly-used techniques to improve software quality~\cite{stroggylos2007refactoring,fowler2018refactoring}. There are different refactoring operations that could be used to improve software quality such as a change in parameter types, move attributes/methods, rename variables/parameters/attributes/methods/classes, extract methods, extract classes, etc \cite{fowler2018refactoring}.

Refactoring plays an important role in software engineering, as its purpose is to improve software quality. Without refactoring, software quality would continue to deteriorate and make development more difficult. Researchers conducted many studies on refactoring in different areas, such as finding the approach to effectively refactor code and determining the impact of refactoring on software quality. One particular aspect of refactoring is increasing the reusability of software components, which provides developers a more efficient way to utilize existing code to create new functionality. Creating reusable software components facilitates development and maintenance since less work is needed to accomplish additional functionality.

While it is usually true that refactoring improves software quality, it is not known how reusability refactoring impacts metrics. Moser et al. \cite{moser2006does} has found that the appropriate refactoring can make the necessary design level changes to improve the software reusability, however, there is no practical evidence on how developers refactor code to improve reusability in practice. 

The purpose of this paper is to investigate how developers use refactoring when they state they are improving code reusability. Therefore, we have mined commits from 1,828 well-engineered project, were we have identified 1,957 reusability commits. We refer to a commit as a \textit{reusability commit} where its developer explicitly mentions, in the commit message, that a refactoring is performed to improve reusability. Then we extract all refactorings executed in these reusability commits, and we label them as \textit{reusability refactorings}. To better understand how developers perceive reusability and apply it in real-world scenarios, we examine how these refactorings manifest in the code by examining their impact on code quality. Furthermore, to check if there are some refactoring patterns that are specific to reusability, we report the distribution of reusability refactorings compared to other refactorings and the distribution of the different types of refactored code elements in reusability refactorings. 
This paper extends a quantitatively and qualitatively our previous study \cite{alomar2020developers}. This papers analyzes the impact of reusability refactorings at a wider set of structural metrics, allowing a better profiling of how the intent of improving design, impacts, either positively or negatively, various design level metrics. From Qualitative point of view, we manually investigate to categorize the intents behind refactoring reusable code. We particularly investigate what triggers developer to refactor the code for the purpose of code reuse.
 To perform this analysis, we formulate the following research questions:

\textbf{RQ1.} \textit{Do developers refactor code differently for the purpose of improving reusability?}

To answer this research question, we execute  Refactoring Miner \cite{tsantalis2018accurate} to extract the type of refactorings that are chosen by developers to improve reusability. We also investigate if there are any refactoring patterns that are specific to reusability, by comparing the distribution of reusability-related refactorings, with the distribution of refactorings for other mainstream development tasks. Then, we identify any significant differences between the distribution values in the two populations.

\textbf{RQ2.} \textit{What is the impact of reusability refactorings on structural metrics?}

To answer this research question, we consider the state-of-the-art reusability structural metrics, extracted from previous studies \cite{moser2006does,alshayeb2009empirical}. We calculate these metrics on files before and after they were refactored for improving reusability. Then we analyze the impact of refactorings on the variation of these metrics, to see if they were capturing the improvement.

\textbf{RQ3.} \textit{
What triggers developers to refactor the code for the purpose of code reuse?}

To answer this research question, we perform case studies that demonstrate GitHub developers’ intentions when refactoring source code to improve code reusability.

The results of our study indicate that when developers make reusability changes, they seem to significantly impact metrics related to methods and attributes, but not parameters or interfaces. Additionally, developers perform reusability changes much less than regular refactoring changes.   Aid from our empirical analysis, we provide the software reuse community with a replication package, containing the dataset we crawled, the files containing all the metric values, for the purpose of replication and extension\footnote{\url{https://smilevo.github.io/self-affirmed-refactoring/}}.

The remainder of this paper is organized as follows: 
 Section \ref{sec:related work} includes some existing studies related to our work. Section \ref{sec:methodology} presents the design of
our empirical study, Section \ref{sec:results} shows the results of our experiments, Section \ref{sec:threat} describes the threats the validity to our study and any mitigation we took to minimize those threats, and Section \ref{sec:conclusion} summarizes the contributions and results of our study.

\section{Related Work}\label{sec:related work}

\textcolor{black}{It is widely acknowledged in the literature of software refactoring that it has the ultimate goal to improve software quality and fix design and implementation bad practices \cite{fowler2018refactoring}. As shown in Table \ref{Table:CK and other metrics}, there is much research effort have focused on studying and exploring the impact of refactoring on software quality \cite{moser2007case,wilking2007empirical,alshayeb2009empirical,shatnawi2011empirical,bavota2015experimental,chavez2017does,mkaouer2016use,cedrim2016does,hegedHus2010effect,alomar2019impact,hamdi2021empirical}. The vast majority of studies have focused on measuring the internal and external quality attributes to determine the overall quality of a software system being refactored. In this section, we review and discuss the relevant literature on software quality in general and reusability in particular.}

\begin{table*}
  \centering
	 \caption{\textcolor{black}{A summary of the literature on the impact of refactoring activities on quality.}}
	 \label{Table:CK and other metrics}
\begin{tabular}{|l|l|l|l|}\hline
\textbf{Study} & \textbf{Year} & \textbf{Approach} & \textbf{Software Metric}  \\  \hline
Sahraoui et al. \cite{sahraoui2000can} & 2000 & Analyzing code histories & CLD / NOC / NMO / NMI       \\ 
& & & NMA / SIX / CBO / DAC   \\
& & & IH-ICP / OCAIC / DMMEC / OMMEC   \\ \hline
Stroulia \& Kapoor \cite{stroulia2001metrics} & 2001 & Performing a case study & LOC / LCOM / CC \\ \hline
Tahvildari et al. \cite{tahvildari2003quality} & 2003 & Analyzing code histories & LOC / CC / CMT / Halstead's efforts    \\ \hline
Leitch \& Stroulia \cite{leitch2003assessing}& 2003 & Analyzing code histories & SLOC / No. of Procedure  \\ \hline
Bois \& Mens \cite{du2003describing} & 2003 & Analyzing code histories & NOM / CC / NOC / CBO   \\ 
& & & RFC / LCOM  \\ \hline
Tahvildari \& Kontogiannis \cite{tahvildari2003metric} & 2004 & Analyzing code histories & LCOM / WMC / RFC / NOM  \\ 
& & & CDE / DAC / TCC \\ \hline
Moser et al. \cite{moser2006does} & 2006 & Analyzing code histories & CK / MCC / LOC    \\ \hline
Wilking et al. \cite{wilking2007empirical} & 2007 & Analyzing code histories & CC / LOC    \\ \hline
Stroggylos \& Spinells \cite{stroggylos2007refactoring} & 2007 & Mining commit log & CK / Ca / NPM  \\ \hline
Moser et al. \cite{moser2007case} & 2008 & Analyzing code histories & CK / LOC / Effort (hour)  \\ \hline
Alshayeb \cite{alshayeb2009empirical} & 2009 & Analyzing code histories &  CK / LOC / FANOUT    \\ \hline
Hegedus et al. \cite{hegedHus2010effect} & 2010 & Analyzing code histories & CK    \\ \hline
Shatnawi \& Li \cite{shatnawi2011empirical} & 2011 & Analyzing code histories & CK / QMOOD   \\  \hline
Bavota et al. \cite{bavota2013empirical} & 2013 & Analyzing code histories & ICP / IC-CD / CCBC  \\
& & Surveying developers & \\ \hline
Szoke et al. \cite{szoke2014bulk} & 2014 & Mining commit log & CC / U / NOA / NII / NAni \\
& & Surveying developers & LOC / NUMPAR / NMni / NA   \\ \hline
Bavota et al. \cite{bavota2015experimental} & 2015 & Mining commit log &  CK / LOC / NOA / NOO  \\
& & Analyzing code histories & C3 / CCBC \\ \hline
Mkaouer et al. \cite{mkaouer2016use} & 2016 &  Many-objective SBSE & QMOOD \\ \hline
Cedrim at al. \cite{cedrim2016does} & 2016 & Mining commit log & LOC / CBO / NOM / CC   \\
& & Analyzing code histories & FANOUT / FANIN   \\ \hline
Chavez et al. \cite{chavez2017does} & 2017 & Mining commit log & CBO / WMC / DIT / NOC  \\ 
& & Analyzing code histories & LOC / LCOM2 / LCOM3 / WOC  \\
& & & TCC / FANIN / FANOUT / CINT  \\
& & & CDISP / CC / Evg / NPATH    \\
& & & MaxNest / IFANIN / OR / CLOC\\
& & & STMTC / CDL / NIV / NIM / NOPA  \\ \hline 
Pantiuchina et al. \cite{pantiuchina2018improving} & 2018 & Mining commit log & LCOM / CBO / WMC / RFC    \\
&  & Analyzing code histories &  C3 / B\&W / Sread \\ \hline
AlOmar et al. \cite{alomar2019impact} & 2019 & Mining commit log & CK / FANIN / FANOUT / CC / NIV / NIM  \\
& &   Analyzing code histories & Evg / NPath / MaxNest / IFANIN   \\ 
& & & LOC / CLOC / CDL / STMTC  \\ \hline
AlOmar et al. \cite{alomar2020developers} & 2020 &  Mining commit log & CK / CC / LOC    \\
& &   Analyzing code histories & \\ \hline
Hamdi et al. \cite{hamdi2021empirical} & 2021 &  Mining commit log & LCOM / CBO / WMC / RFC \\
& & Analyzing code histories & NOSI / TCC / LCC / LOC  \\
& & &  VQYT / DIT \\ \hline

\end{tabular}
\end{table*}

\subsection{Studies on Software Reusability}

\textcolor{black}{Software reusability has been a topic of interest since the 1970s, because of that, a large amount of literature that discusses it is available. This section, although not accounting for all available work, does its best in presenting multiple papers from the different ways reusability is discussed. Previous work discussed general aspects of reusability, identifying challenges, topics, issues and principles \cite{Ahmaro,reuseGeneral4,reuseGeneral5,reuseGeneral2,reuseGeneral3,reuseGeneral1,Younoussi}.}

\textcolor{black}{Ahmaro et al. \cite{Ahmaro} conducted a systematic literature review to identify the definition, approaches, benefits, reusability levels, factors, and adoption of software reusability. The authors found that the concept of software reusability consisted of 11 approaches, namely, design patterns, component-based development, application frameworks, legacy system wrapping, service-oriented systems, application product lines, COTS integration, program libraries, program generators, aspect-oriented software development and configurable vertical applications. A study on the relationship of complexity and reuse
design principles is reported by Anguswamy and Frakes \cite{reuseGeneral4}. Their findings show that the higher the
complexity the lower the ease of reuse. Lubars et al. \cite{reuseGeneral2} contrasted code reusability in the large versus code reusability in small with regards to several aspects, including, size, complexity, application, and problems associated with locating and reusing the code. The author highlighted that code reusability in the small has
had limited impact because of its strongly self-centered
orientation, whereas code reusability in the large has had limited impact because of its high degree of difficulty in
finding the reusable components. In a similar context, Mockus \cite{reuseGeneral3} performed large-scale code reuse study in open source software and found that more than 50\% of the files were used in more than one project. Yin and Lee \cite{reuseGeneral1} conducted a survey to examine the characteristics of software reusability from the points of
view of software engineering as well as knowledge engineering. Younoussi and Roudies \cite{Younoussi} presented a systematic literature review on software reusability. They pointed out that few studies examined barriers of reusability, and organizations need to adapt software reusability approaches.}

\textcolor{black}{Reusability and code reuse are also discussed in relation to specific topics \cite{reuseSpecific4,reuseSpecific3,reuseSpecific1,reuseSpecific2,patrick2020exploring,lotter2018code,an2017stack,abdalkareem2017code,alomar2020exploratory,feitosa2020code}. Lotter et al. \cite{lotter2018code} explored code reuse between Stack Overflow and Java open-source systems in order to understand how the practice of reusing code could affect future software maintenance and the correct use of license. Their findings show that there is up to 3.3 \% code reuse within Stack Overflow, while 1.0 \% of Stack Overflow code is reused in Java projects. Patrick \cite{patrick2020exploring} investigated reusability metrics with Q\&A forum data. The author proposed an approach (LANLAN), using word embeddings and machine learning, to classify Q\&A forum posts into support requests and problem reports, as well as reveal information in relation to software reusability and explore potential reusability metrics. In another context, Abdalkareem et al. \cite{abdalkareem2017code} performed an exploratory study on 22 Android apps to explore how much, why, when, and who reuses code. They found that 1.3 \% of the Apps were constructed from Stack Overflow posts, and discovered that mid-aged and older apps reuse Stack Overflow code later in their lifetime. An et al. \cite{an2017stack} also explored Android apps and found that 15.5 \% of the apps contained exact code clones, and 60 out of 62 apps, had potential license violations. Recently, AlOmar et al. \cite{alomar2020exploratory} presented insights regarding how developers discuss software reuse by analyzing Stack Overflow. These findings can be used to guide future research and to assess the relevance  of software reuse nowadays. 
 Their findings show that software reuse is a decreasing trend in Stack Overflow which might indicate that developers have widely adopted this practice and thus few questions regarding it emerge as it is well grasped by the community. Further, Feitosa et al. \cite{feitosa2020code} studied the relation between software reuse at the class level and technical debt. The authors found that reused classes tend to concentrate more principally, and reused code usually has less technical debt interest.}




\subsection{Studies on Software Quality}

\begin{table*}[h!]
\caption{Summary of related studies on developer perception and quality.}
\label{tab:summary-related-studies-tbl}
\begin{adjustbox}{width=1.0\textwidth,center}
\begin{tabular}{|l|l|l|l|l|l|}
\hline
\textbf{Study}              & \textbf{Year} & \textbf{Focus}                                                                                                                                                                                                           & \textbf{Dataset Size}  & \textbf{Quality Attribute} & \textbf{Software Metric}                                                                                                                                                                                                      \\ \hline
Moser et al. \cite{moser2006does}      & 2006 & \begin{tabular}[c]{@{}l@{}}Reusability measurement over time.\end{tabular} &  \begin{tabular}[c]{@{}l@{}}30 Java classes\end{tabular} & \begin{tabular}[c]{@{}l@{}}Reusability\end{tabular} & \begin{tabular}[c]{@{}l@{}} LCOM / RFC / CC \\ CBO / WMC /LOC \\ DIT /NOC\end{tabular}                                                                                                         \\ \hline
Pantiuchina et al. \cite{pantiuchina2018improving} & 2018 & \begin{tabular}[c]{@{}l@{}}Developer's perception \& quality\end{tabular}                                                                                                         & \begin{tabular}[c]{@{}l@{}}\\ 1,282 commits \end{tabular} & \begin{tabular}[c]{@{}l@{}} Cohesion / Coupling \\ Complexity / Readability \end{tabular} & \begin{tabular}[c]{@{}l@{}}LOCM / C3 / CBO \\ RFC / WMC / B\&W \\ Sread\end{tabular}   \\ \hline
Fakhoury et al. \cite{fakhoury2019improving}    & 2019 & \begin{tabular}[c]{@{}l@{}}Developer's perception \& quality \end{tabular}                                                       & \begin{tabular}[c]{@{}l@{}}548 commits\end{tabular} & \begin{tabular}[c]{@{}l@{}} Readability\end{tabular} & \begin{tabular}[c]{@{}l@{}}B\&W / Sread / Dorn\end{tabular}            \\ \hline
AlOmar et al. \cite{alomar2019impact}      & 2019 & \begin{tabular}[c]{@{}l@{}} Developer's perception \& quality \end{tabular}  & \begin{tabular}[c]{@{}l@{}}1,245 commits\end{tabular} & \begin{tabular}[c]{@{}l@{}} Coupling / Cohesion \\ Complexity / Inheritance \\ Polymorphism / Encapsulation \\ Abstraction / Size \end{tabular} & \begin{tabular}[c]{@{}l@{}} LCOM / CBO / FANIN \\ FANOUT / RFC /CC \\ WMC / Evg / NPATH \\MaxNest / DIT / NOC \\ IFANIN / LOC / CLOC \\ STMTC / CDL / NIV \\ NIM  \end{tabular}                                                                                                          \\ \hline
AlOmar et al. \cite{alomar2020developers}    & 2020 & \begin{tabular}[c]{@{}l@{}} Developer's perception \& quality\end{tabular}  & \begin{tabular}[c]{@{}l@{}}1,967 commits\end{tabular} & \begin{tabular}[c]{@{}l@{}} Reusability \end{tabular} & \begin{tabular}[c]{@{}l@{}} LCOM / CBO / RFC \\ CC / WMC / LOC \\ DIT / NOC  \end{tabular}                                                                                                          \\ \hline
\end{tabular}
\end{adjustbox}
\end{table*}

Research in refactoring software has covered a variety of aspects, including tools and methods to facilitate refactoring and accurately assess the impact of refactoring on software quality. 
Pantiuchina et al. \cite{pantiuchina2018improving} talked about determining if there was a difference in how developers perceive refactorings will be helpful, and how the metrics say the refactorings were. That study determined that even if a developer reports that there was a refactoring done it might not be reflected in the metrics. This study focuses on comparing specific refactorings relating to certain metrics, specifically \say{cohesion}, \say{coupling}, \say{readability}, and \say{complexity}, to metrics that measure those attributes, while we focused on using metrics to determine if there was a quantifiable difference, and if so, what that difference was, during self-proclaimed reusability refactorings. Even then, something to take away from this study is that measuring refactoring code changes focusing on quality of life, rather than strictly functional, can have many moving parts not measured by metrics. Metrics do not tell the whole story, and while it is good to see what metrics are affected when developers improve reusability, it could also be helpful to include information and narratives from actual developers alongside the pure metrics.

Fakhoury et al. \cite{fakhoury2019improving} have shown that the existing readability models are not able to capture the readability improvement with minor changes in the code, and some metrics which can effectively measure the readability improvement are currently not used by readability models. The authors also studied the distribution of different types of changes in readability improvements, which is similar to our research question, which examines the distribution of the different types of refactored code elements in reusability refactorings. 

Prior works \cite{alomar2019can,peruma2020contextualizing,alomar2020toward} have explored how developers document their refactoring activities in commit messages; this activity is called Self-Admitted Refactoring or Self-Affirmed Refactoring (SAR). In particular, SAR indicates developers' explicit documentation of refactoring operations intentionally introduced during a code change.

AlOmar et al. \cite{alomar2019impact} showed that there is a misperception between the state-of-the-art structural metrics widely used as indicators for refactoring and what developers consider to be an improvement in their source code. The research aims to identify (among software quality models) the metrics that align with the vision of developers on the quality attribute they explicitly state they want to improve. Their approach entailed mining 322,479 commits from 3,795 open source projects, from which they identified about 1,245 commits based on commit messages that explicitly informed the refactoring towards improving quality attributes. Thereafter, they processed the identified commits by measuring structural metrics before and after the changes. The variations in values were then compared to distinguish metrics that are significantly impacted by the refactoring, towards better reflecting the intention of developers to improve the corresponding quality attribute. Our study also utilized software quality metrics to evaluate the impact of refactoring on reusability. 

Research particularly in reusability refactoring by Moser et al. \cite{moser2006does} showed that refactoring increases the quality and reusability of classes in an industrial, agile environment. Similar to our paper, their study examines the impact of refactoring on quality metrics related to reusability on the method and class levels, such as Weighted Method per Class (WMC) and Coupling Between Object (CBO), respectively. The results of their experiment revealed that refactoring significantly improved the metrics Response for Class (RFC) and Coupling Between Object classes (CBO) related to reusability. However, the limitations of their study involved a small project consisting of 30 Java classes and 1,770 Lines of Code (LOC) developed by two pairs of programmers over the course of 8 weeks. In addition, the authors considered how general refactoring operations impact metrics related to reusability, rather than specifically reusability refactorings. In our study, we examined 1,828 projects and 154,820 commits that modified Java files. We also considered how reusability changes affect software quality metrics and how what kinds of refactoring operations were performed during reusability changes. Table \ref{tab:summary-related-studies-tbl} shows the summary of each study related to our work.



\textcolor{black}{Our work highlights on the aspect of reusability refactorings, and is different from the above-mentioned studies as our main purpose is to explore if there is an alignment between quality metrics and reusability quality improvement that are documented by developers in the commit messages. To the best of our knowledge, no previous study has empirically investigated, using a curated set of commits, and the representativeness of
structural design metrics for reusability quality attribute.}


\section{Experimental Design}\label{sec:methodology}

\begin{figure*}[t]
\centering 
\includegraphics[width=1.0\textwidth]{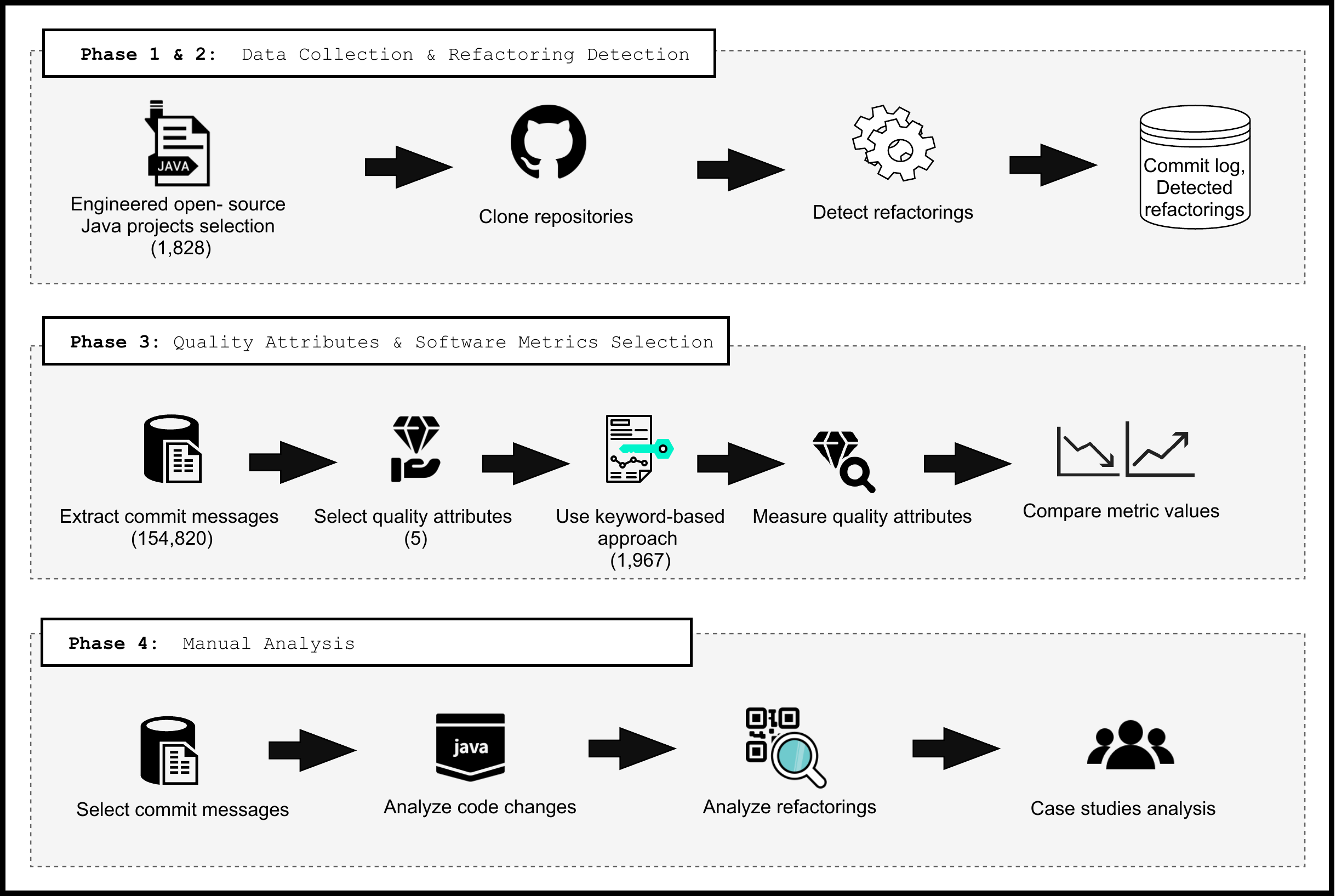}
\caption{Empirical study design overview.}
\label{fig:approach_overview}
\end{figure*}

\textcolor{black}{According to the guidelines reported by Runeson and H{\"o}st \cite{runeson2009guidelines}, we design an empirical study that is supported by explanatory case studies \cite{robson2002real}. Our research method consists of three steps as depicted in Figure \ref{fig:approach_overview}}.  We detail each activity of our methodology in the subsequent subsections. The dataset utilized in this study is available for extension and replication purpose \footnote{\url{https://smilevo.github.io/self-affirmed-refactoring/}}.

\subsection{Selection of Quality Attributes and Structural Metrics}
\textcolor{black}{We started by conducting a literature review on existing and well-known software quality metrics and their corresponding possible measurements \cite{chidamber1994metrics,lorenz1994object,mccabe1976complexity}. Next, we extracted metrics that are used to assess several object-oriented design aspects in general, and software reusability in particular. For example, the RFC (Response for Class) metric is typically used to measure visibility of a given class in the project, the more a class is responsive, the more it can be accessed and its functionality can be reused by other objects in the system. More generally, we extract, from literature review, all the associations between metrics (e.g., CK
suite \cite{chidamber1994metrics}, McCabe \cite{mccabe1976complexity}) with reusability quality attribute.}
 
\textcolor{black}{The process left us with 8 object-oriented metrics as shown in Table \ref{table:reusability_metrics}. The list of metrics is (1) well-known and defined in the literature, and (2) can assess on different code-level elements, i.e., method, class, package, and (3) can be calculated by the tool we considered. All metrics values are automatically computed using the tool \textsc{Understand}\footnote{\url{https://scitools.com/}}, a software quality assurance framework.}

\begin{table*}
\centering
\caption{Reusability and its corresponding
structural metrics used in this study.}\label{table:reusability_metrics}
\begin{tabular}{|l|l|l|}
\hline
\textbf{Quality Attribute} &  \textbf{Study} & \textbf{Software Metric}\\
\hline
Cohesion & \cite{alshayeb2009empirical,moser2006does} & Lack of Cohesion of Methods (LCOM) \\ \hline
Complexity & \cite{moser2006does} & Response for Class (RFC) \\ 
& \cite{moser2006does} & Cyclomatic Complexity (CC) \\  
& \cite{chavez2017does,alomar2019impact} & Paths (NPATH) \\ 
& \cite{chavez2017does,alomar2019impact} & Nesting (MaxNest) \\ \hline
Coupling & \cite{alshayeb2009empirical,moser2006does} & Coupling Between Objects (CBO) \\ 
& \cite{chavez2017does,alomar2019impact} & Fan-in (FANIN) \\
& \cite{chavez2017does,alomar2019impact} & Fan-out (FANOUT) \\ \hline 
Design Size & \cite{alshayeb2009empirical,moser2006does} & Weighted Method per Class (WMC)\\ 
& \cite{alshayeb2009empirical,moser2006does} & Line of Code (LOC) \\ 
& \cite{chavez2017does,alomar2019impact} & Lines with Comments (CLOC) \\
& \cite{chavez2017does,alomar2019impact} & Statements (STMTC) \\
& \cite{chavez2017does,alomar2019impact} & Classes (CDL) \\
& \cite{chavez2017does,alomar2019impact} & Instance Variables (NIV) \\
& \cite{chavez2017does,alomar2019impact} & Instance Methods (NIM) \\ \hline
Inheritance & \cite{alshayeb2009empirical,moser2006does} & Depth of Inheritance Tree (DIT) \\
& \cite{alshayeb2009empirical,moser2006does} & Number of Children (NOC) \\ 
& \cite{chavez2017does,alomar2019impact} & Base Classes (IFANIN) \\ \hline
\end{tabular}
\end{table*}

\subsection{Refactoring Detection}
The projects in our study consist of 1,828 open-source Java projects, which were curated projects hosted on GitHub. These projects were selected from a dataset made available by Munaiah et al.  \cite{munaiah2017curating}, while verifying that these are Java-based projects since this is the only language the Refactoring Miner \cite{tsantalis2018accurate} supports. These projects utilize software engineering practices such as documentation and testing. 

\textcolor{black}{We utilize Refactoring Miner \cite{tsantalis2018accurate} for mining refactorings from each project in our dataset. Refactoring Miner is designed to analyze code changes (i.e., commits) in Git repositories to detect applied refactorings. Our choice of the mining tool is driven by its accuracy (precision of 98\% and a recall of 87\%) and is suitable for a study that requires a high degree of automation since it can be used through its external API.}

In this phase, we collect a total of 862,888 refactoring operations from 154,820 commits. An overview of the studied benchmark is provided in Table~\ref{Table:DATA_Overview}.

\begin{table}[h]
\begin{center}
\caption{Studied dataset statistics.}
\label{Table:DATA_Overview}
\begin{tabular}{|l|l|}
\hline
\textbf{Item} &  \textbf{Count} \\
\hline
Studied projects & 1,828 \\
Commits with refactorings & 154,820 \\
Commits with \textit{reus*/reusability} Keywords & 1,967 \\
Reusability refactoring operations & 3,065 \\
\hline
\end{tabular}
\end{center}
\end{table}

\subsection{Reusability Commits Extraction}

After extracting all refactoring commit messages detected by Refactoring Miner, our next step consists of analyzing each of the commit messages as we want to only keep commits where refactoring is documented, i.e., self-affirmed refactoring (SAR) \cite{alomar2019can,alomar2020we,alomar2020toward}. As for the commit message selection, we initially use a keyword-based approach to find those commits that contain the keywords \textit{reus*}\footnote{Regular expression was used to capture all expansions of reus such as reuses, reusing, reuse, etc.} and  \textit{reusability}. We have chosen these two keywords because of their popularity in the development community as being used by developers to describe software reusability \cite{sharma2007critical}. We then kept commits whose messages contained the two keywords. \textcolor{black}{We performed a manual analysis of all the commits, and we ended up removing any duplicates and false positives. This was done by the first two authors.} This process resulted in selecting 1,967 commits, containing 3,065 refactorings, as our dataset for this study. Each dataset instance is a commit, along with its corresponding refactorings.

\begin{figure}[ht]
\centering 
\includegraphics[width=1.0\columnwidth]{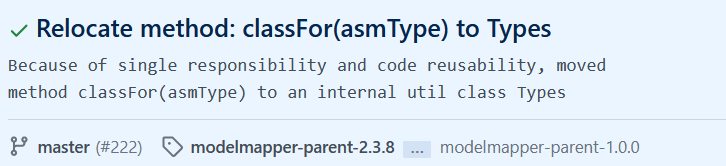}
\caption{A sample instance of our dataset.}
\label{fig:example}
\end{figure}

As an illustrative example, Figure \ref{fig:example} details a commit whose message states the relocation of the method \textit{classFor(asmType)} to an internal class utility class for the purpose of applying the single responsibility principle and code reusability\footnote{https://github.com/modelmapper/modelmapper/commit/ 6796071fc6ad98150b6faf654c8200164f977aa4}. After running Refactoring Miner, we detected the existence of a \textit{Move method} refactoring from the class \textit{ExplicitMappingVisitor} to the class \textit{Types}. The detected refactoring matches the description of the commit message, and gives more insights about the old placement of the method, which was absent in the textual description. As we explain in the following subsection, we need to locate all the code elements involved in the refactoring (source class, target class, etc.) for the purpose of evaluating the quality of the relocation in terms of impact of structural metrics, such as coupling and cohesion.




\subsection{Metrics Calculation}

To generate the metric values for reusability commits, we ran code evaluation tool, specifically using \textsc{Understand}\footnote{https://scitools.com/features/}. The metrics we used to evaluate the code quality are summarized in Table \ref{table:reusability_metrics}.

We then used SQL queries to find reusability commits in the dataset and their associated project links to clone using Git and exported the results from our dataset to a combined Comma-Separated Value (CSV) file. Using a shell script, we cloned the projects, checked out the versions for each commit, and ran the Git diff command to see which files changed in each commit. \textcolor{black}{The goal of using Git diff is to track the changed files in each commit and categorize them into the 'before/after' directory for further analysis. For each file being modified in a commit, the Git diff shows the file paths and a status related to the change as follows:}
\begin{itemize}
    \item \textcolor{black}{'M' (modification of the contents or mode of a file)}
    \item \textcolor{black}{'R' (renaming of a file)}
    \item \textcolor{black}{'D' (deletion of a file)}
    \item \textcolor{black}{'A' (addition of a file)}
\end{itemize}

\textcolor{black}{Based on the four different status, we performed different actions on the file.}
\begin{itemize}
    \item \textcolor{black}{For ‘M’, we keep the original file before the change in ‘before’ directory, and the modified file after the change in ‘after’ directory.}
    \item \textcolor{black}{For ‘R’, we keep the original file before the change, and the renamed file after the change in ‘after’ directory.} 
\item \textcolor{black}{For ‘D’, we keep the file in the ‘before’ directory.}
\item \textcolor{black}{For ‘A’, we keep the file in the ‘after’ directory.}

\end{itemize}

In another words, if files were deleted in a commit, we included the metric values for those files before the commit but not after it. If files were added in a commit, we included the metric values for those files after the commit but not before it. If files were renamed or moved in a commit, then we included the metric values for those files both before and after the commit. Our shell script then ran the \textsc{Understand} tool to generate metrics for the changed files for the versions before and after each reusability commit, resulting in two files containing metric values for each commit: (1) one file for the files changed before the commit and (2) another file for the files changed after the commit. 

Since each metric value before and after the commit are dependent to each other, we decided to use the Wilcoxon Signed-rank Test \cite{wilcoxon1945individual} to determine whether or not there were statistically significant differences in the metric values for all changed files before and after the reusability commits.  \textcolor{black}{We formulated the following hypotheses}:

\textcolor{black}{\(H_0\): \textit{There was no improvement in the metrics we analyzed between before and after the reusability refactoring}.} 


\textcolor{black}{\(H_1\): \textit{There was an improvement shown as an increase}}. 

To achieve that, we created Python scripts to order and sort all the values from the above results from \textsc{Understand} to ensure that the rows in both before and after files are corresponding to each other. 
 Next, we combined the data in the CSV files before and after the commits together into another two CSV files each have a total of 185,244 metric values: one CSV file for all code elements in changed files before the reusability commits, and another CSV file for all code elements in changes files after the commits. 
 The Wilcoxon Signed-rank Test allowed us to determine if any metrics were statistically significantly changed when developers performed self-proclaimed reusability refactorings. 
 
\subsection{Manual Analysis}

\textcolor{black}{To get a more qualitative sense of the context of code reuse involving the act of refactoring, we create case studies that demonstrate GitHub developers’ intentions when refactoring
source code for the purpose of code reuse. Case study is one of the empirical methods used for studying phenomena in a real-life context \cite{wohlin2000experimentation}}. \textcolor{black}{In our study, we performed
a combination of manual analysis and quantitative analysis. For each case study, we checkout the corresponding commit to obtain the source code, then two authors manually analyze the code changes. We provide the commit message and its corresponding refactoring operations detected by the tool Refactoring Miner}. We elaborate in detail these case studies in Section \ref{RQ3}, where we report on our results.

\section{Results}\label{sec:results}

This section reports and discusses our experimental results and aims to answer our research questions. 

\subsection{\textbf{RQ1.} Do developers refactor code differently for the purpose of improving reusability?}

This research question aims to compare refactoring activity in reusability commits with the refactoring activity that can be found in mainstream development tasks (feature updates, bug fix, etc.). Since we have a dataset of all refactorings performed in the 1,828 projects that we study, we separate refactorings that belong to the reusability commits (refactorings performed for the purpose of improving reusability), which we refer to as \textit{reusability refactorings}. We refer to the remaining refactorings as \textit{non-reusability refactorings}. Then, for each group, we calculate the percentage of each refactoring type, among the total refactorings of that group.

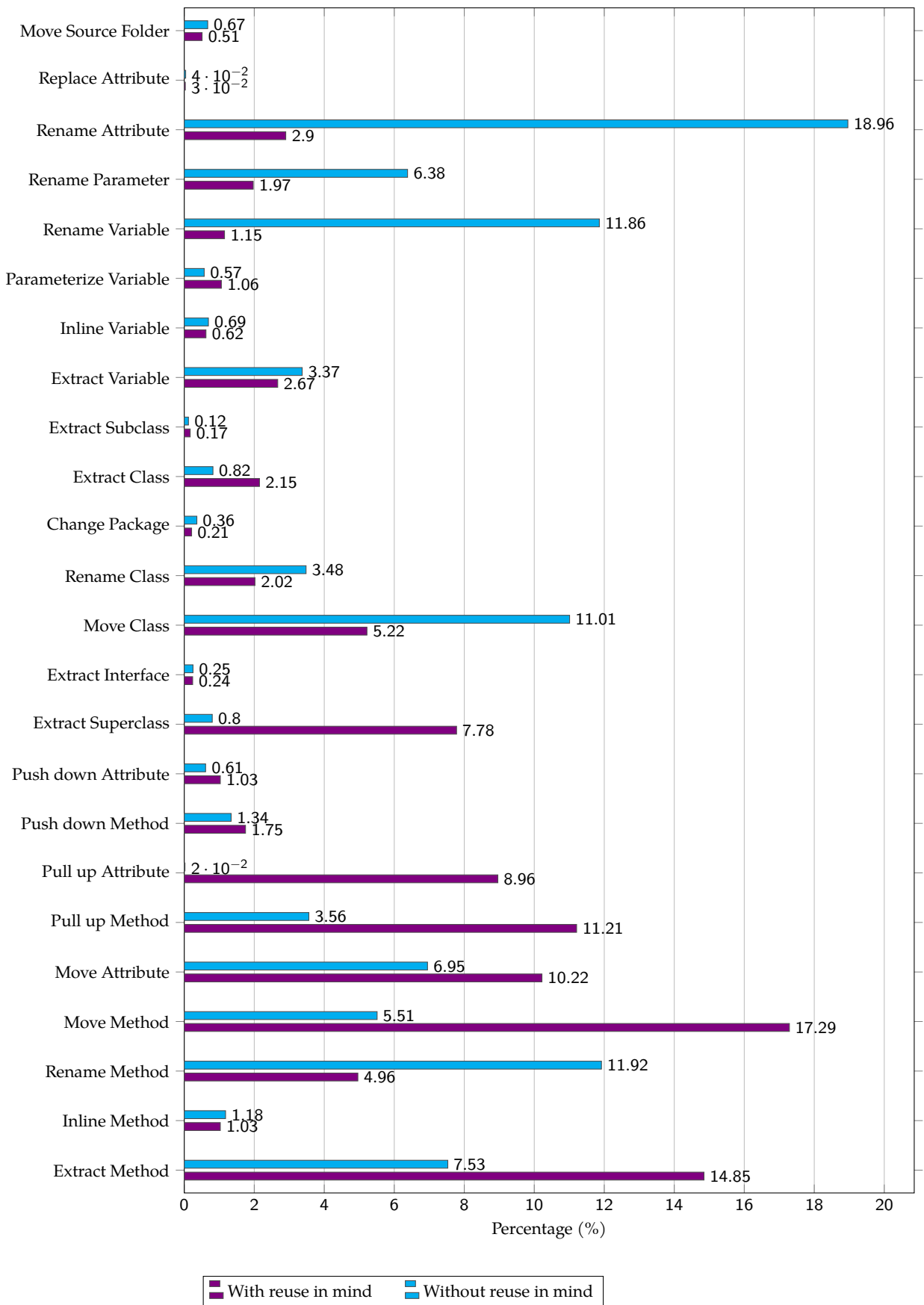
\begin{figure*}[htbp]
\centering
\begin{tikzpicture}
\pgfplotsset{%
    width=.83\textwidth,
    height=1.3\textwidth
}
\begin{axis}[ 
    xmajorgrids=true,
    xbar, 
    xmin=0,
    xlabel={Percentage (\%)},
    symbolic y coords={{Extract Method}, {Inline Method}, {Rename Method},{Move Method}, {Move Attribute}, {Pull up Method}, {Pull up Attribute}, {Push down Method}, {Push down Attribute},{Extract Superclass}, {Extract Interface}, {Move Class}, {Rename Class},{Change Package}, {Extract Class},{Extract Subclass}, {Extract Variable}, {Inline Variable}, {Parameterize Variable}, {Rename Variable},{Rename Parameter},{Rename Attribute},{Replace Attribute},{Move Source Folder}},
    ytick=data,
    legend style={
    	at={(0.3,-0.07)},
        anchor=north,
        legend columns=-1,
        /tikz/every even column/.append style={column sep=0.5cm}
        },
    nodes near coords,
    every node near coord/.append style={font=\small},
    nodes near coords align={horizontal},
     enlarge y limits=0.02,
]
\addplot[fill=violet,draw=black!70,tickwidth = 0pt,bar width=4pt,label style={font=\small}, tick label style={font=\small}] coordinates {(14.85,{Extract Method}) (1.03,{Inline Method}) (4.96,{Rename Method}) (17.29,{Move Method}) (10.22,{Move Attribute}) (11.21,{Pull up Method}) (8.96,{Pull up Attribute}) (1.75,{Push down Method}) (1.03,{Push down Attribute}) (7.78,{Extract Superclass}) (0.24,{Extract Interface}) (5.22,{Move Class}) (2.02,{Rename Class}) (0.21,{Change Package}) (2.15,{Extract Class}) (0.17,{Extract Subclass}) (2.67,{Extract Variable}) (0.62,{Inline Variable}) (1.06,{Parameterize Variable}) (1.15,{Rename Variable}) (1.97,{Rename Parameter}) (2.9,{Rename Attribute}) (0.03,{Replace Attribute}) (0.51,{Move Source Folder})};
\addplot[fill=cyan,draw=black!70,tickwidth = 0pt,bar width=4pt,label style={font=\small}, tick label style={font=\small}] coordinates {(7.53,{Extract Method}) (1.18,{Inline Method}) (11.92,{Rename Method}) (5.51,{Move Method}) (6.95,{Move Attribute}) (3.56,{Pull up Method}) (0.02,{Pull up Attribute}) (1.34,{Push down Method}) (0.61,{Push down Attribute}) (0.8,{Extract Superclass}) (0.25,{Extract Interface}) (11.01,{Move Class}) (3.48,{Rename Class}) (0.36,{Change Package}) (0.82,{Extract Class}) (0.12,{Extract Subclass}) (3.37,{Extract Variable}) (0.69,{Inline Variable}) (0.57,{Parameterize Variable}) (11.86,{Rename Variable}) (6.38,{Rename Parameter}) (18.96,{Rename Attribute}) (0.04,{Replace Attribute}) (0.67,{Move Source Folder})};
\legend{\textcolor{black}{With reuse in mind}, \textcolor{black}{Without reuse in mind}}
\end{axis}
\end{tikzpicture}
\caption{Percentages of \textit{reusability refactoring} and \textit{non-reusability refactorings}, clustered by type.}
\label{fig:distirbution}
\end{figure*}

Figure \ref{fig:distirbution} visualizes, by percentage of the total refactoring operations in each of the respective sets, the distributions of refactoring operations. We observe that the distribution of \textit{reusability refactorings} varies from the \textit{non-reusability refactorings}. In fact, the top frequent types in reusability refactorings are, \textit{Move Method}, \textit{Extract Method}, and \textit{Pull-Up Method}, whose percentages are respectively, 17.29\%, 14.85\%, and 11.21\%. For non-reusability refactorings, the top frequent type were \textit{Rename Attribute}, \textit{Rename Method}, and \textit{Rename Variable}, as their percentages are respectively, 18.96\%, 11.92\%, and 11.86\%. While the \textit{move} related types were highly solicited in reusability refactorings, the \textit{rename} activity was dominant for non-reusability refactorings, which was expected since previous studies who analyzed mainstream refactoring has found that renames are the most popular refactorings \cite{tsantalis2018accurate,alomar2019impact,peruma2020contextualizing,peruma2020exploratory}. However, reusability refactorings seem to be different. To analyze the extent to which reusability and non-reusability refactorings vary, we compare the distribution of refactoring refactorings identified for each group using the Wilcoxon signed-rank test, a pairwise statistical test verifying whether two sets have a similar distribution \cite{wilcoxon1945individual}. If the p-value is smaller than 0.05, the distribution difference between the two sets is considered statistically significant. The choice of Wilcoxon comes from its non-parametric nature with no assumption of a normal data distribution. Upon running the statistical test, the null hypothesis was rejected and the difference between group distributions was found to be statistically significant. 

\textcolor{black}{Another interesting observation that we draw is the popularity of method-level refactoring, being in  TOP 3 most frequent reusability refactorings. Figure \ref{fig:codeelement} shows the distribution of code elements impacted by refactorings, and we notice that more than 50\% of refactorings were performed at the method level.}

To better understand the observed results, we sampled a subset of reusability refactorings, and we have extracted two main patterns:
\begin{figure*}[h]
\centering 
\begin{tikzpicture}
\begin{scope}[scale=0.85]
\pie[rotate = 180,pos ={0,0},text=inside,outside under=30,no number]{50.48/Method\and50.12\%,20.28/Attribute\and20.14\%, 13.43/Class\and13.33\%, 0.24/Interface\and0.24\%,11.57/Variable\and11.57\%,3.99/Parameter\and3.99\% }
\end{scope}
\end{tikzpicture}
\caption{Distribution of code elements in reusability refactoring commits.}
\label{fig:codeelement}
\end{figure*}
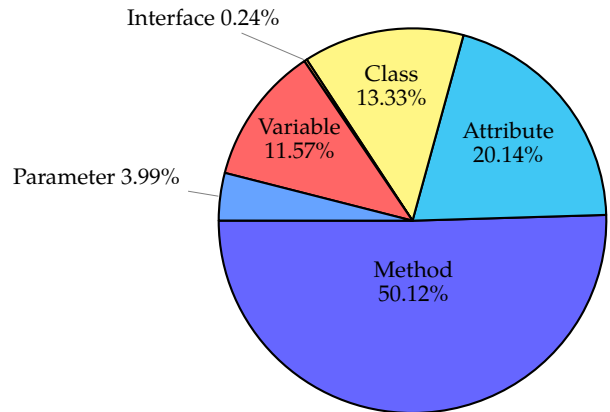

\textbf{Functionality extraction.} When developers are interested in a needed functionality, which is found inside a long method, containing various functionalities, they extract the code elements, belonging to the needed functionality, into a newly created separate method, and they update the original method with the appropriate method calls. This decomposition process is known as \textit{Extract Method}. The newly extracted method has its own visibility, which is independent from the original method, and so developers can increase its visibility of the purpose of reuse, and so other objects and methods can now access it.

\textbf{Functionality movement.} To increase the reusability of a given method, we have noticed that developers typically move methods from less visible classes, into more visible classes, in the system. Various methods were moved into utility classes, which are eventually offering their services to the other classes in the system, this explains why \textit{Move Method} was the most popular type in reusability refactorings, according to Figure \ref{fig:distirbution}. Our qualitative analysis has also shown scenarios of moving method up, from a child class, into a super class, for the purpose of sharing its behavior across all subclasses through inheritance. This refactoring is known as \textit{Pull-Up Method}, which was found to be the third popular type in reusability refactorings, while being not popular in non-reusability refactorings. 







\begin{tcolorbox}
\textbf{Summary.}
We have shown that the distribution of refactoring types, applied in the context of reusability, is different from the distribution of refactoring types in mainstream development. In the refactorings performed to improve reusability, files are subject to more design level types of refactorings (e.g., \textit{Move Method}, \textit{Extract Method}) in general, and inheritance-related refactorings (e.g., \textit{Pull-up Method}, \textit{Pull-up Attribute}) in particular, while in other refactorings, files tend to undergo more renames (e.g., \textit{Rename Method}, \textit{Rename Variable}) and data type changes (e.g., \textit{Change Variable Type}) to identifiers. Reusability refactorings heavily impact, high-level code elements, such as packages, classes, and methods, while typical refactorings, impact all code elements, including identifiers, and parameters. 
\end{tcolorbox}

\subsection{\textbf{RQ2. What is the impact of reusability refactorings on structural metrics?}}

\begin{figure*}[htbp]
\centering
\begin{subfigure}{6cm}
\centering\includegraphics[width=6cm]{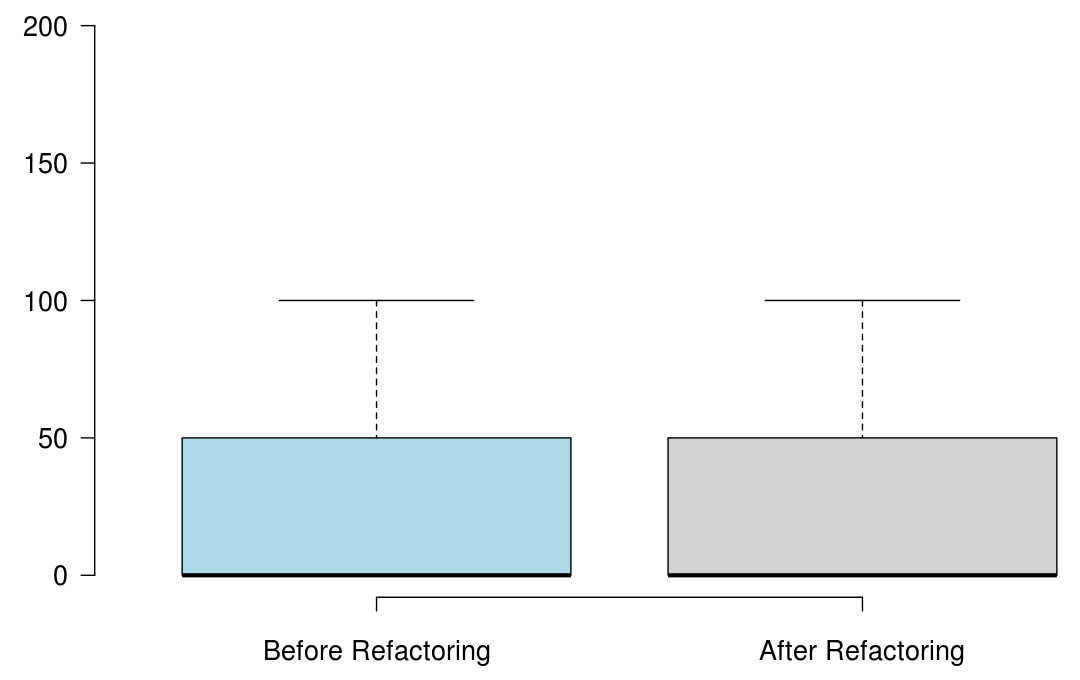}
\caption{Percent Lack of Cohesion}
\label{BP:lcom}
\end{subfigure}%
\begin{subfigure}{6cm}
\centering\includegraphics[width=6cm]{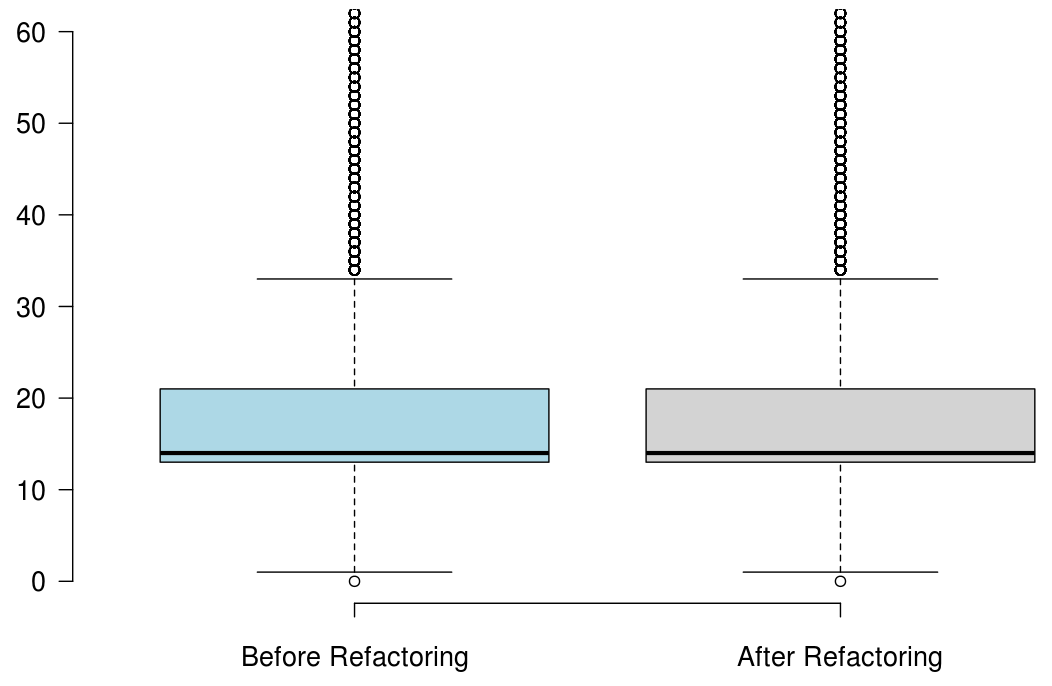}
\caption{Response for Class}
\label{BP:rfc}
\end{subfigure}

\begin{subfigure}{6cm}
\centering\includegraphics[width=6cm]{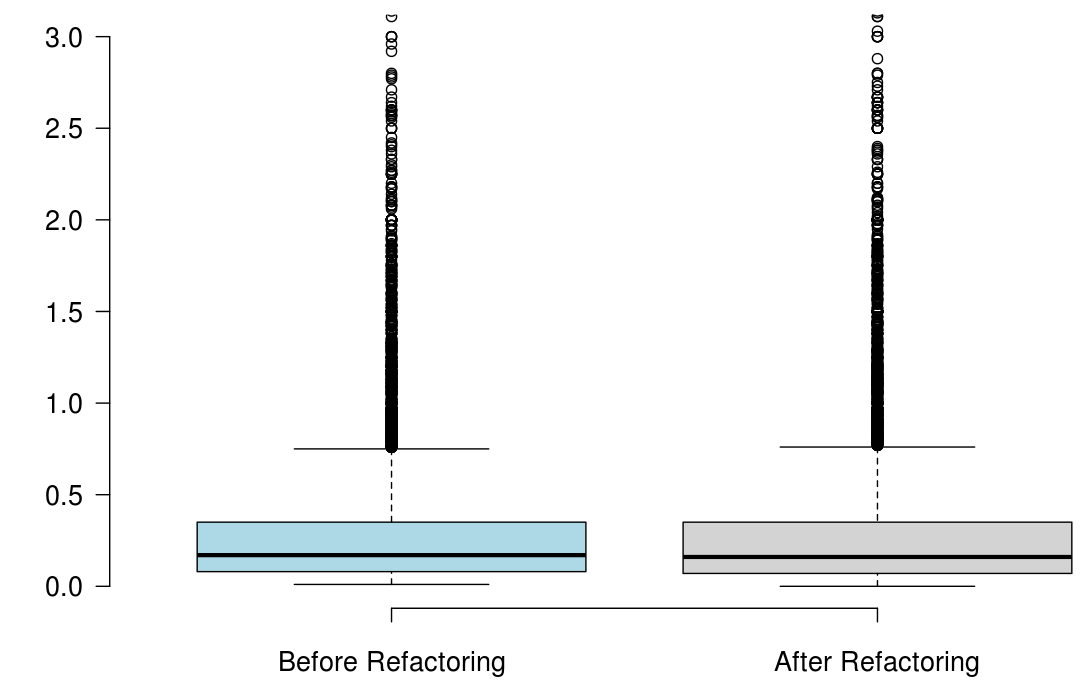}
\caption{Cyclomatic Complexity}
\label{BP:cc}
\end{subfigure}
\begin{subfigure}{6cm}
\centering\includegraphics[width=6cm]{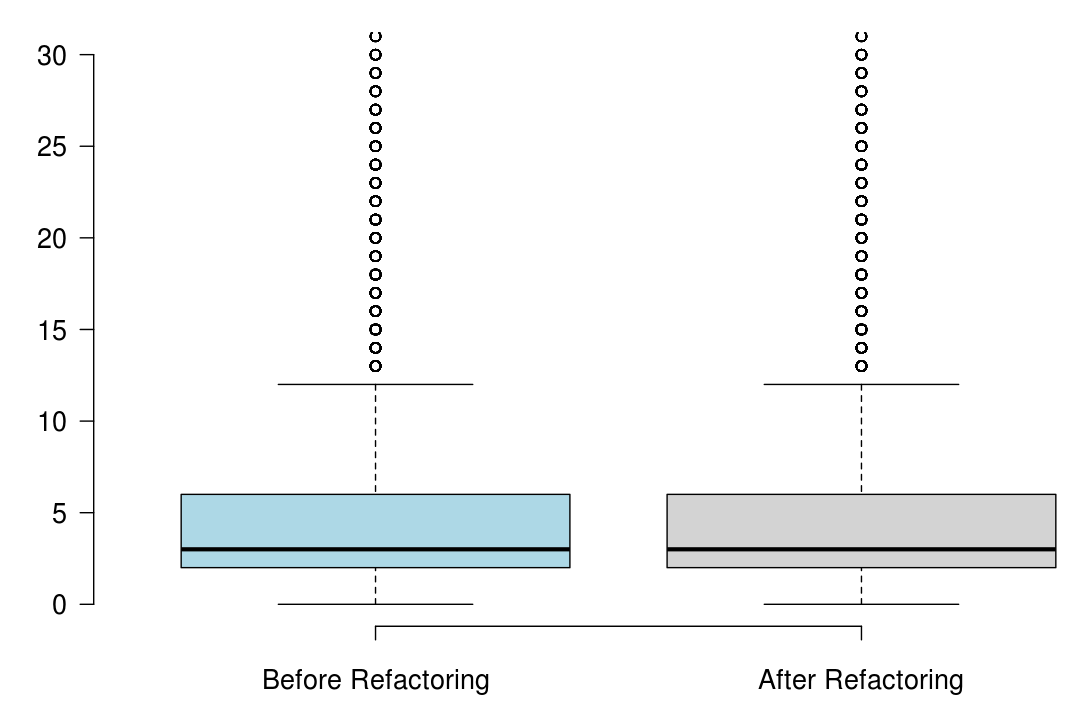}
\caption{Coupling Between Object}
\label{BP:cbo}
\end{subfigure}

\begin{subfigure}{6cm}
\centering\includegraphics[width=6cm]{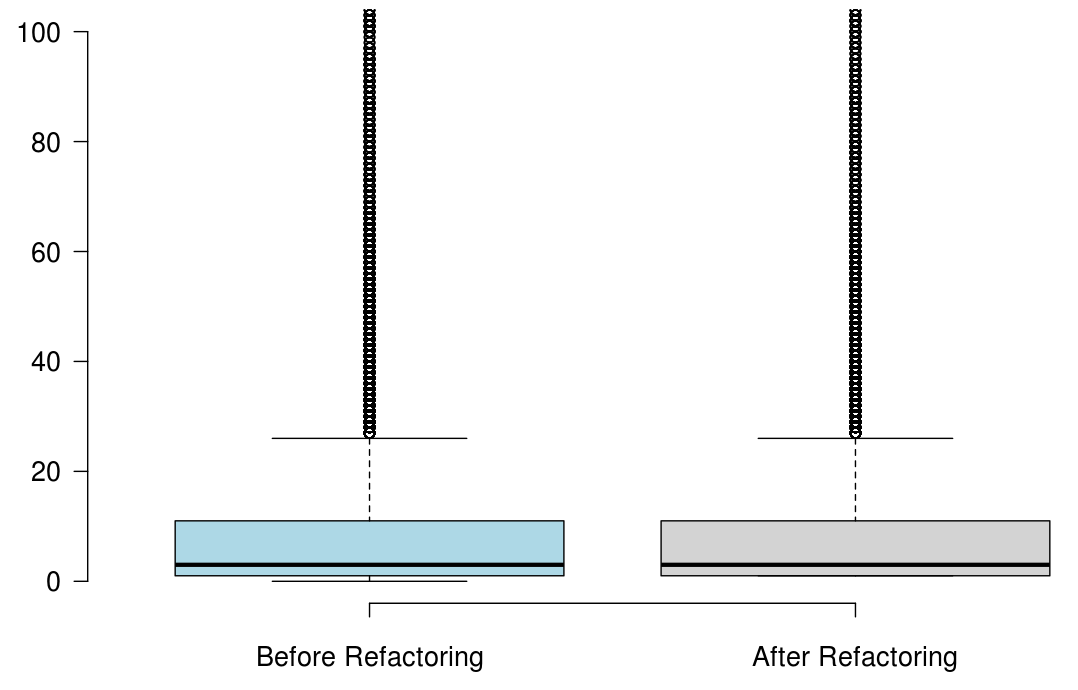}
\caption{Weighted Method per Class}
\label{BP:wmc}
\end{subfigure}%
\begin{subfigure}{6cm}
\centering\includegraphics[width=6cm]{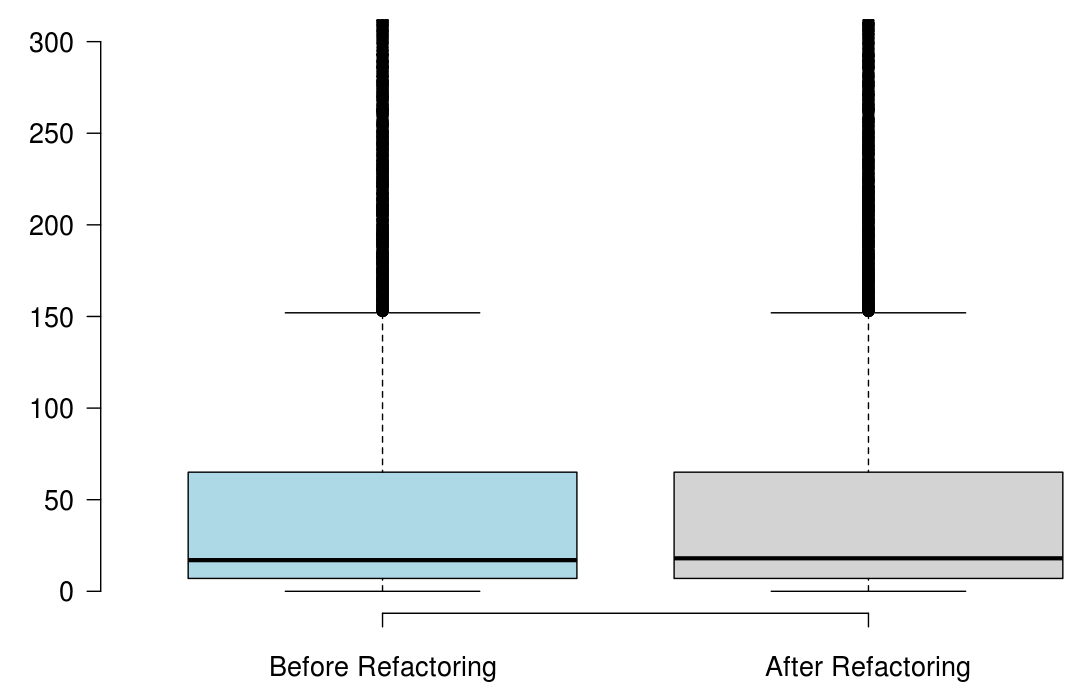}
\caption{Line of Code}
\label{BP:loc}
\end{subfigure}%

\begin{subfigure}{6cm}
\centering\includegraphics[width=6cm]{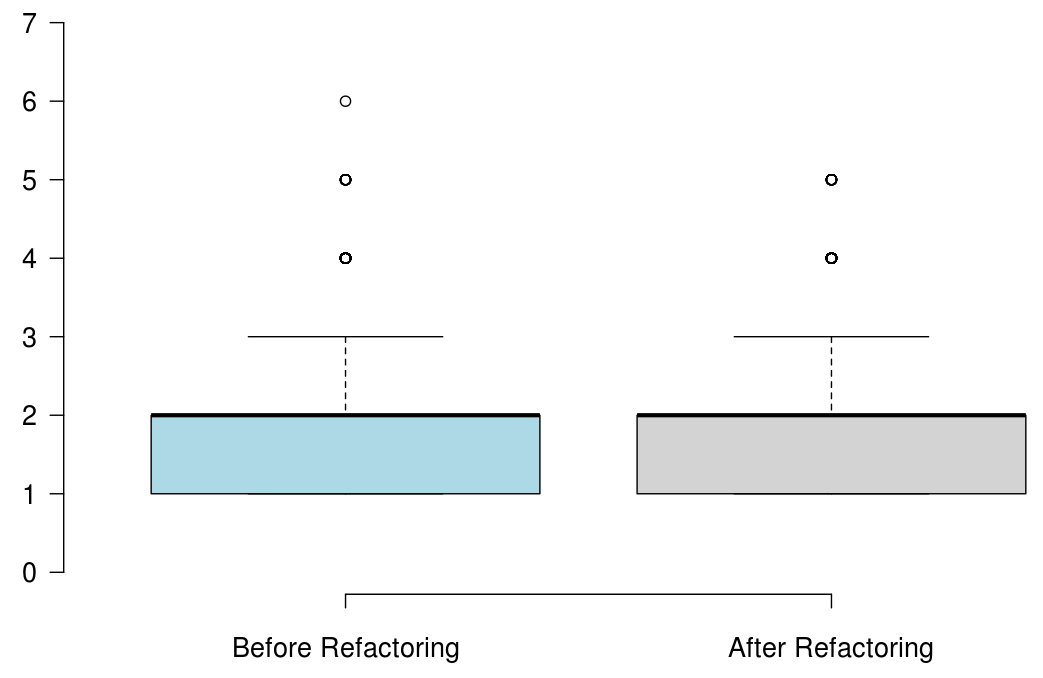}
\caption{Depth of Inheritance Tree}
\label{BP:dit}
\end{subfigure}%
\begin{subfigure}{6cm}
\centering\includegraphics[width=6cm]{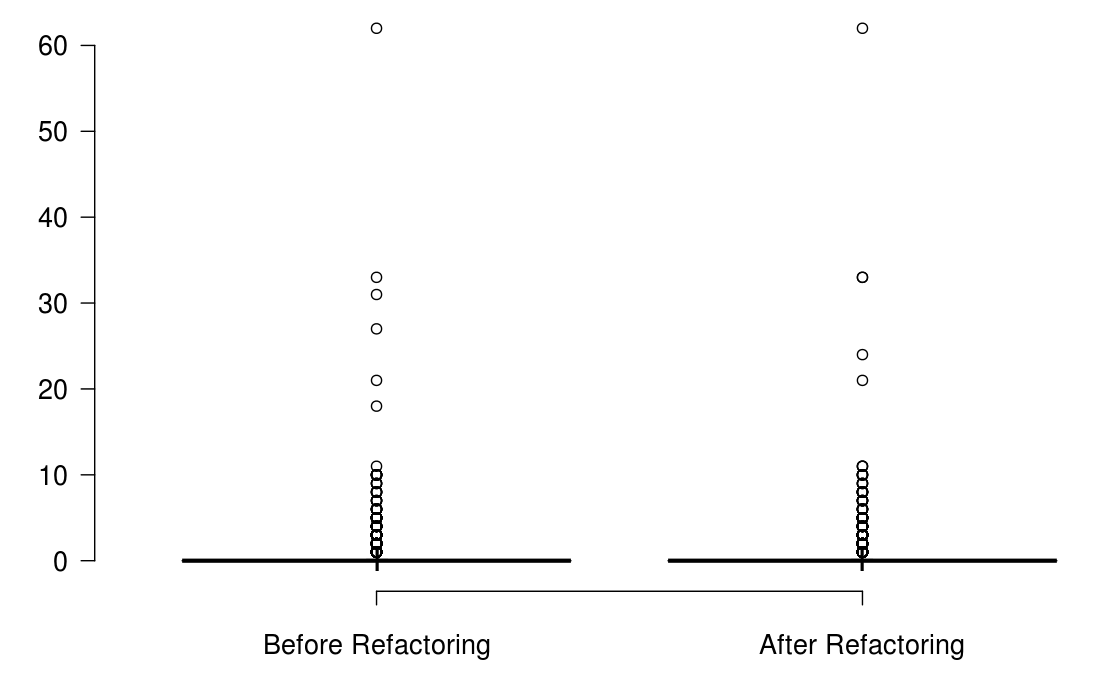}
\caption{Number of Children}
\label{BP:noc}
\end{subfigure}%

\caption{Boxplots for metric values before and after reusability commits for different sets of code elements.} 
\label{fig:rq1-metric-box-plots}
\end{figure*}

To answer this research question, we investigate the impact of reusability refactorings on the  state-of-the-art metrics, which have been used by previous studies, to recommend reusability changes. As a reminder, we aim to look at the variation of each metric value after the execution of the refactoring, therefore, we checkout the project files, right before the reusability commit, and we calculate metrics values, and after the reusability commit, and we recalculate the metrics values. Note that we only consider files that were involved in the commit, as there files are considered part of developer's intention of improving reusability. The results of metrics boxplots are outlined in Figure \ref{fig:rq1-metric-box-plots}. To further investigate the significance of difference between the boxplots, we also use the Wilcoxon Signed-rank Test. Statistical settings included using a 0.05 alpha value for the significance level. We hypothesize that reusability refactorings will optimize metrics by reducing them (the lower is the value of the metric, the better is the software structural quality). Our alternative hypothesis is accepted if the \textit{before refactoring} boxplot is significantly larger than the \textit{after refactoring} boxplot. The Wilcoxon Signed-rank Test results indicating whether or not there were statistically significant improvements before and after reusability commits is shown in Table \ref{tab:rq1-stat-test-results}.

\begin{figure*}[htbp]
\centering
\begin{subfigure}{6cm}
\centering\includegraphics[width=6cm]{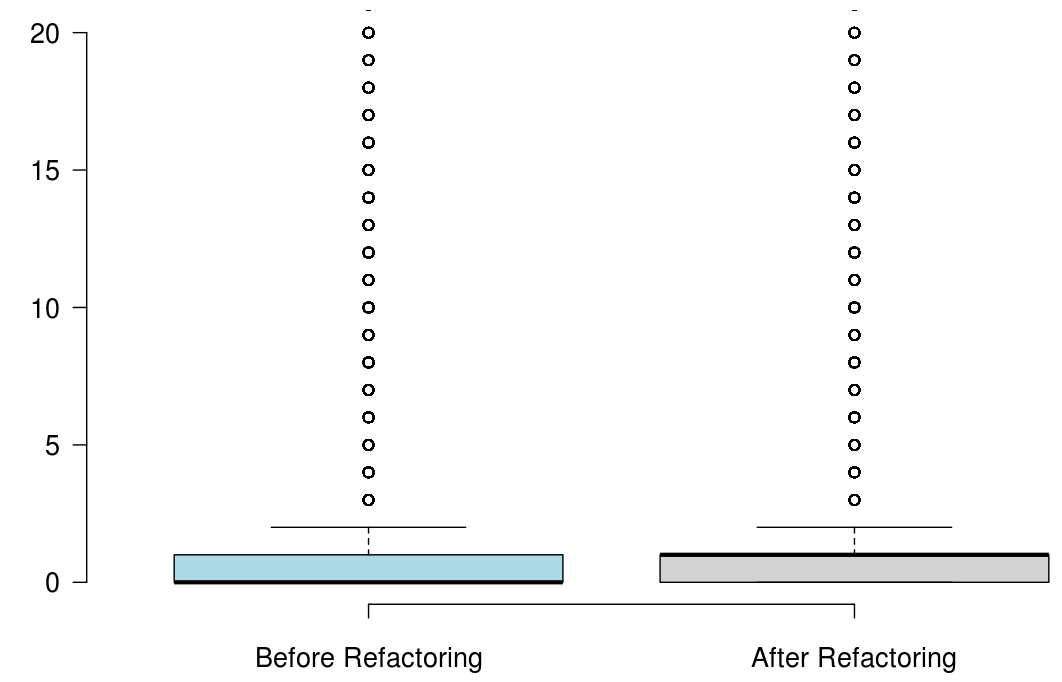}
\caption{Paths}
\label{BP:npath}
\end{subfigure}%
\begin{subfigure}{6cm}
\centering\includegraphics[width=6cm]{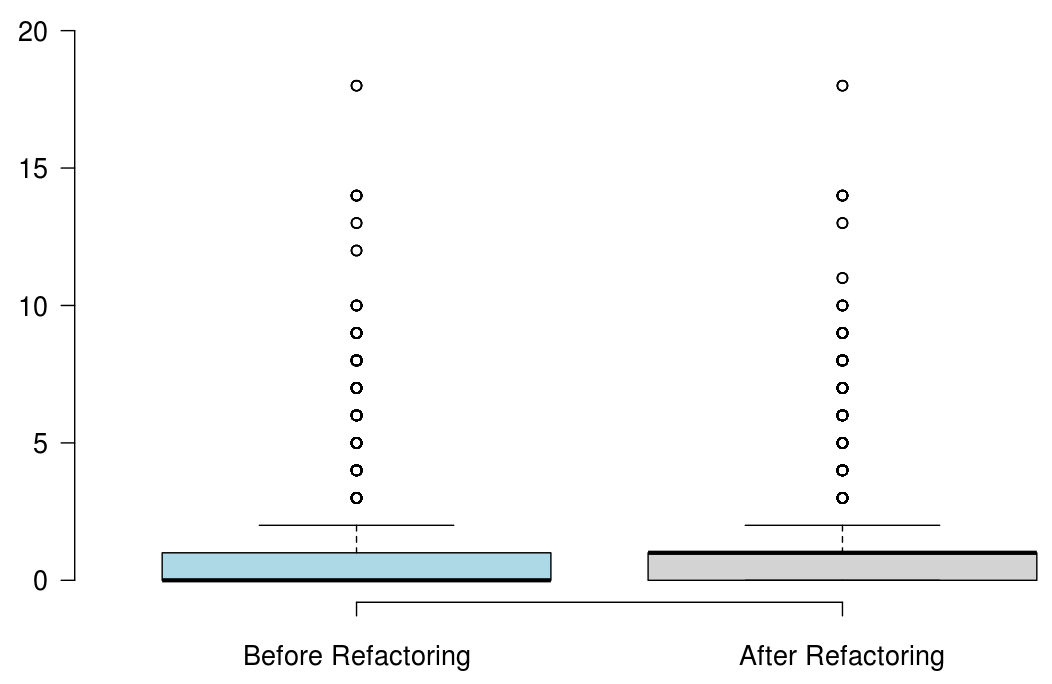}
\caption{Nesting}
\label{BP:maxnest}
\end{subfigure}%

\begin{subfigure}{6cm}
\centering\includegraphics[width=6cm]{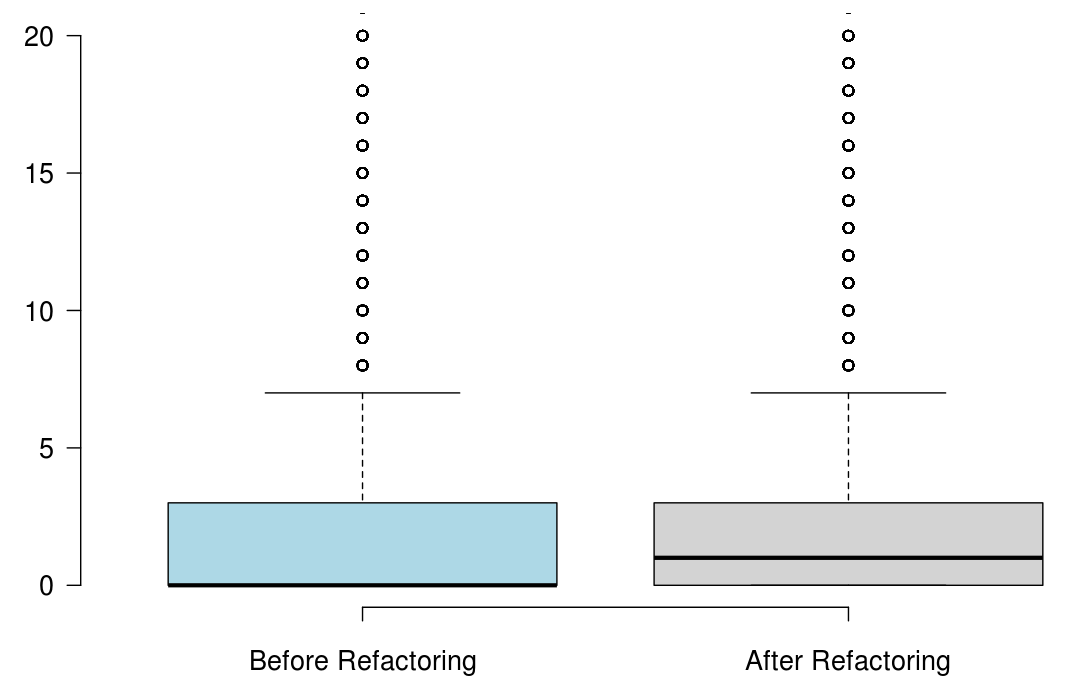}
\caption{Fan-in}
\label{BP:fanin}
\end{subfigure}%
\begin{subfigure}{6cm}
\centering\includegraphics[width=6cm]{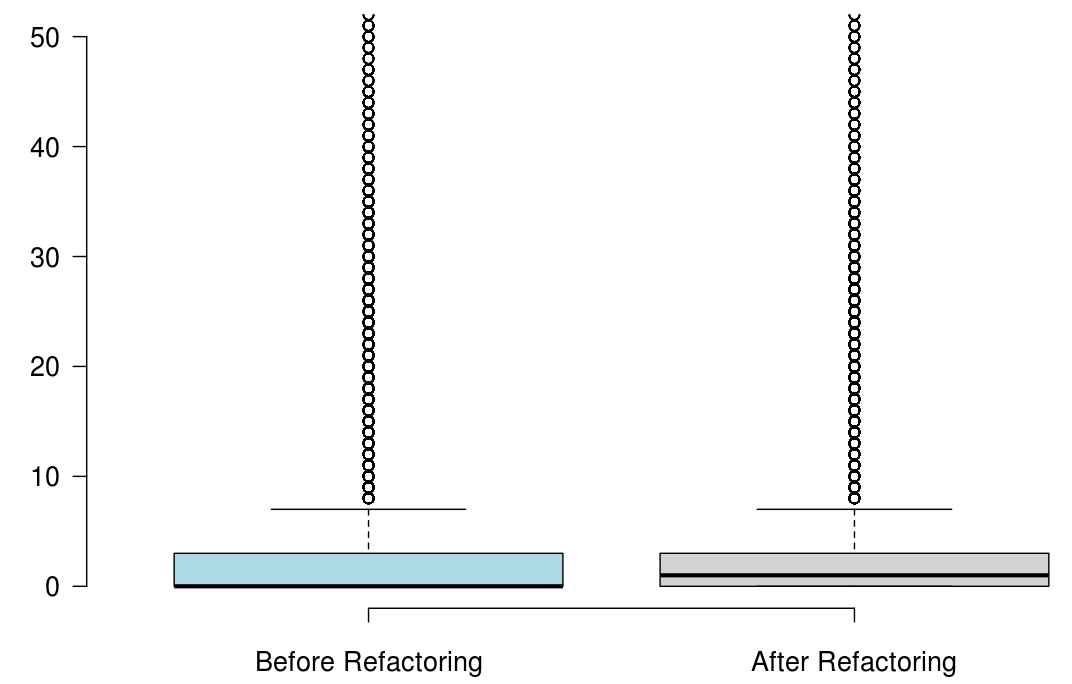}
\caption{Fan-out}
\label{BP:fanout}
\end{subfigure}%

\begin{subfigure}{6cm}
\centering\includegraphics[width=6cm]{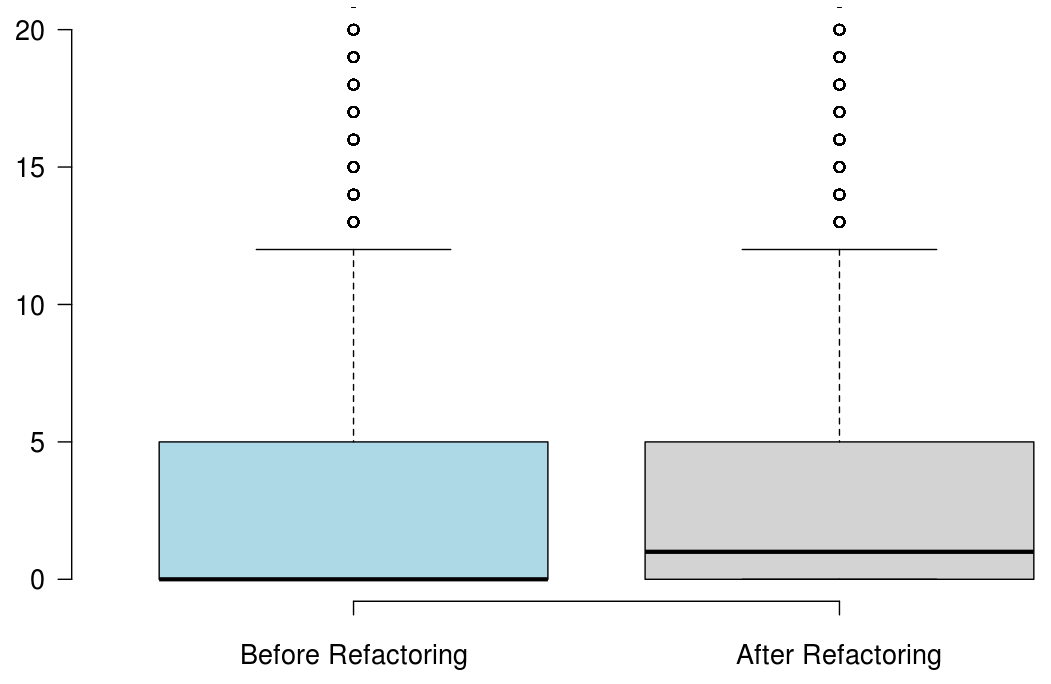}
\caption{Lines with Comments}
\label{BP:cloc}
\end{subfigure}%
\begin{subfigure}{6cm}
\centering\includegraphics[width=6cm]{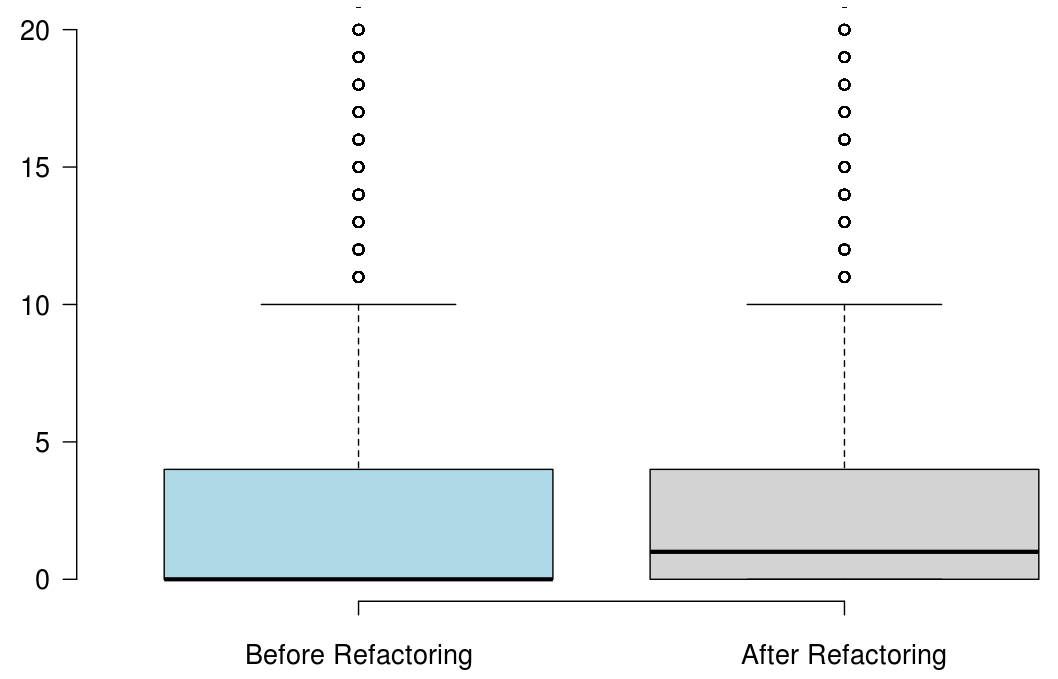}
\caption{Statements}
\label{BP:stmtc}
\end{subfigure}%

\begin{subfigure}{6cm}
\centering\includegraphics[width=6cm]{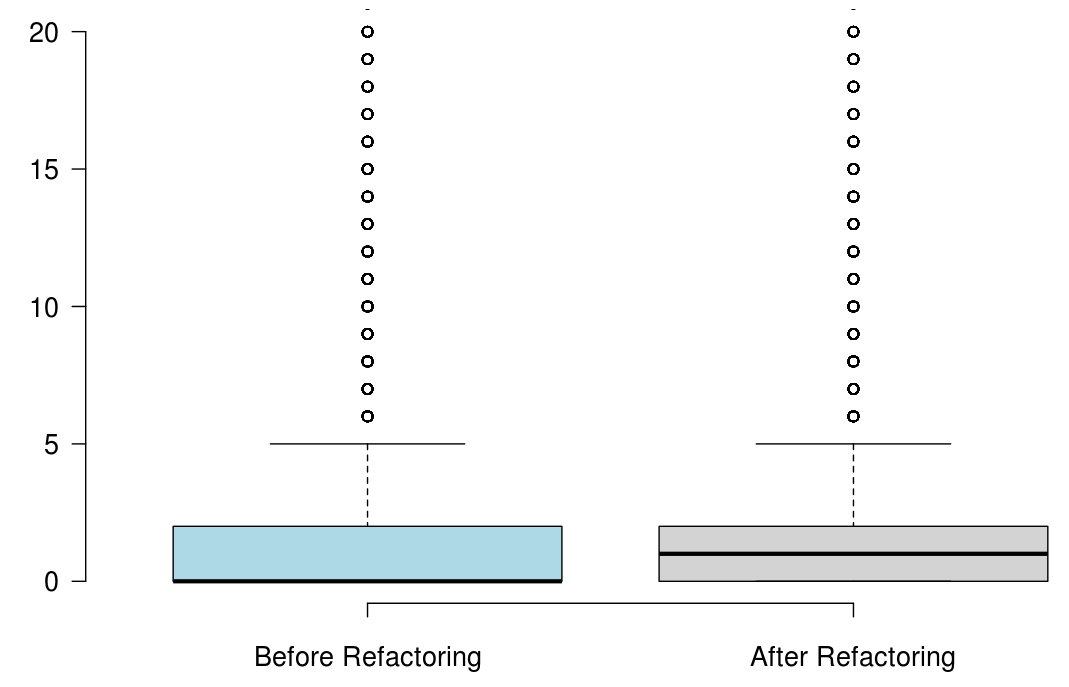}
\caption{Classes}
\label{BP:cdl}
\end{subfigure}%
\begin{subfigure}{6cm}
\centering\includegraphics[width=6cm]{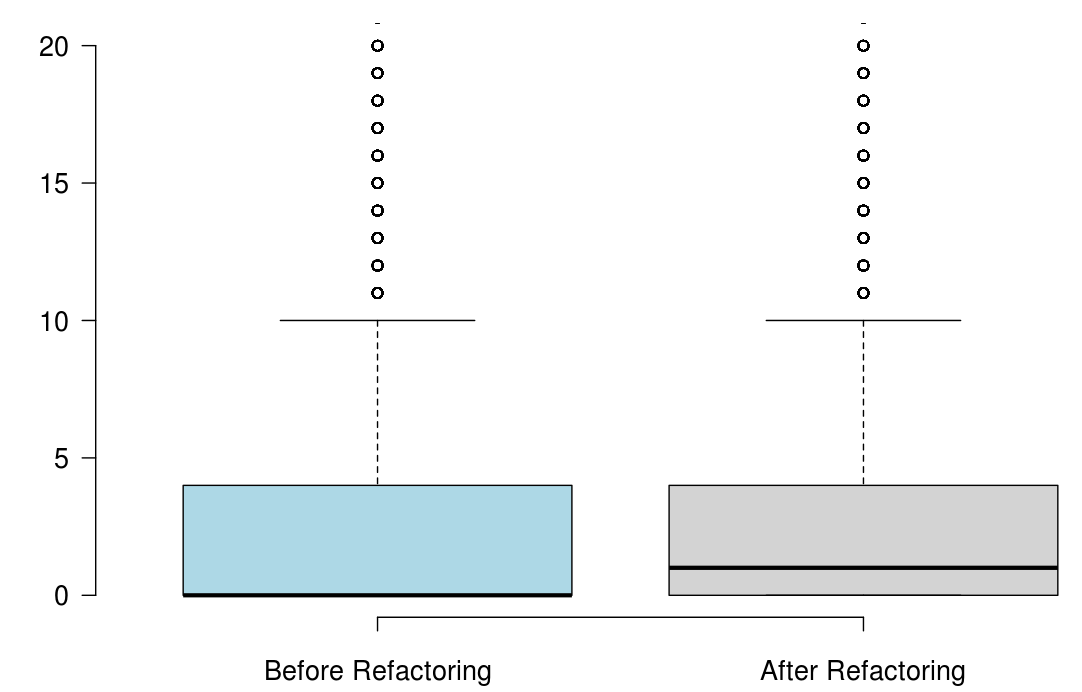}
\caption{Instance Variables}
\label{BP:niv}
\end{subfigure}%

\begin{subfigure}{6cm}
\centering\includegraphics[width=6cm]{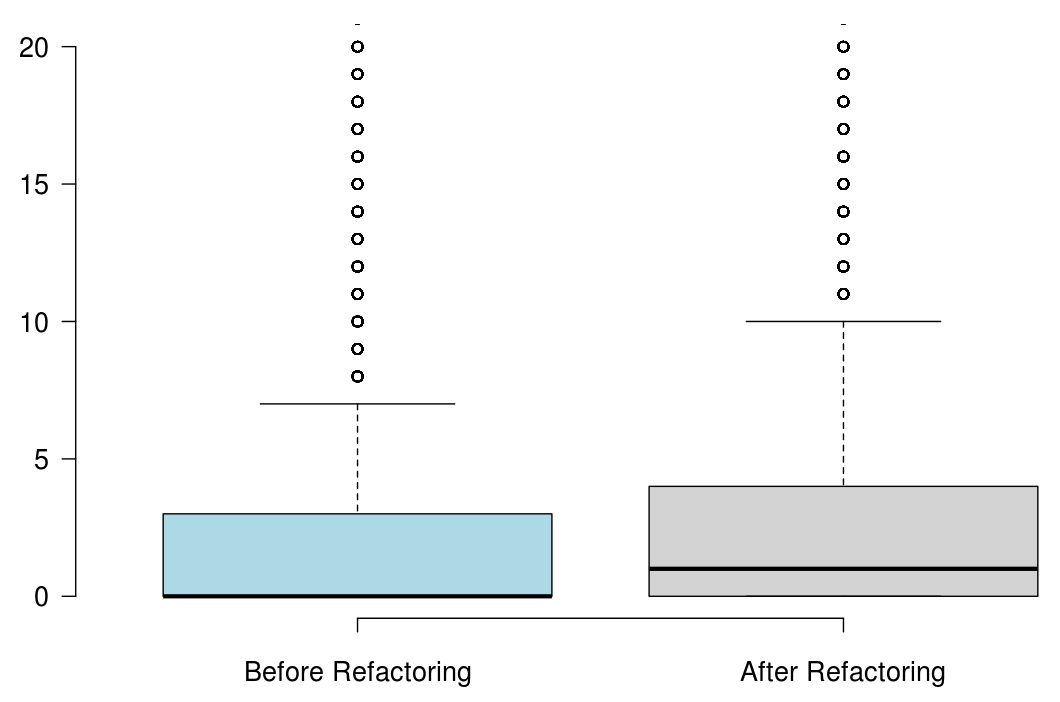}
\caption{Instance Methods}
\label{BP:nim}
\end{subfigure}%
\begin{subfigure}{6cm}
\centering\includegraphics[width=6cm]{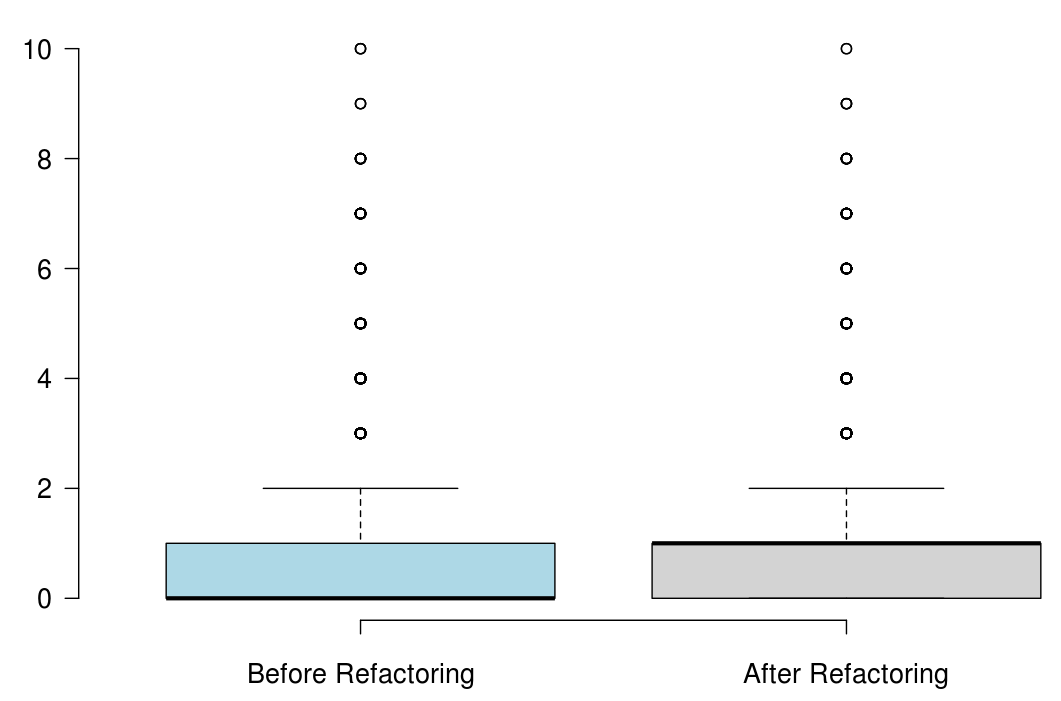}
\caption{Base Classes}
\label{BP:ifanin}
\end{subfigure}%

\caption{
Boxplots for metric values before and after reusability commits for different sets of code elements (Cont.).}
\label{fig:rq1-metric-box-plots_2}
\end{figure*}

According to Figure \ref{fig:rq1-metric-box-plots}, reusability refactorings had no impact on the Number of Children (NOC), Depth of Inheritance Tree (DIT), and Response for Class (RFC). These results can be explained by the fact that the majority of reusability refactoring are not targeting classes. In fact, if we refer to Figure \ref{fig:codeelement}, only 13.3\% of reusability refactoring targeted classes, and exctrating subclasses, which would have impacted these metrics, represent only 0.13\%, and so, its impact is negligible.

On the other hand, we measure an increase in the weighted methods per class, and the variation is found to be statistically significant ($p<0.05$). According to Figure \ref{fig:distirbution}, the \textit{Extract Method} refactoring has been found to be very popular in reusability refactoring, and so, developers tend to create new methods while extracting the reusable code from the longer methods. This implies the sudden increase of methods count, per class. While developers are expected to keep the number of methods lower in classes, the impact of reusable functionality from longer classes, creates free methods that can be pulled up to either superclasses, and be shared with all children, or relocated to operate on variables that may not belong to its original class. This explains decrease of the Coupling Between Objects (CBO) and the slight decrease in the Lack of Cohesion of Methods (LCOM), which means that methods have become more cohesive. However, its corresponding statistical test show no significant different, but its value was close to 0.05. Similarly, we notice slight improvement in the Lines of Code (LOC), with no statistical significance but close p-value (i.e., 0.066). The extraction of methods helps in reducing cloning functionalities in multiple locations in the code. Also, pulling methods up the hierarchy, will allow subclasses to inherit it, and so, lines of code will decrease, unless when the method gets overridden. 
Moreover, the Cyclomatic Complexity (CC) has decreased after reusability code changes with no statistical significance. A proper extraction of sub-methods tends to break down long methods, and slightly decrease their complexity.

As shown in Figure \ref{fig:rq1-metric-box-plots_2}, we measure an increase in the following metrics: the Number of Paths (NPATH), the Fan-in (FANIN), the Fan-out (FANOUT), the Number of Instance Methods (NIM), and the variation is found to be not statistically significant. This observation indicates that developers increase number of possible paths, the number of calling subprograms plus global variables read, the number of called subprograms plus global variables set, and the number of instance methods. We also observe We notice the improvement of fine metrics, namely in the Nesting (MaxNesting), the Lines with Comments (CLOC), the Number of Statements (STMTC), the Number of Classes (CDL), the Number of Instance Variables (NIV), after the commits in which developers explicitly target the improvement
of the reusability refactoring, but with no statistical significance. This indicates that developers reduce the maximum nesting level of control constructs, the line containing comments, the number of statements, the number of classes and the number of declared instance variables. Further, Similar to the findings of the Depth of Inheritance Tree (DIT) and the Number of Children (NOC), reusability refactorings had no impact on the Number of Base Classes (IFANIN), which emphasizes on the fact that most of the reusability refactorings are not targeting immediate base classes.

As a meta-review, the majority of state-of-the-art metrics did not capture any improvement, or captured non-significant improvement, when developers refactor their code for the purpose of reusability. This is an interesting finding for our future research directions, as we want to further increase our dataset, in terms of projects, and programming languages, in order to experiment whether there is a shortage of metrics that properly measure what developers consider to be at design level change to improve reusability. Such investigations will bridge the gap between how existing research on software reuse evaluates code changes, and how developers concretely achieve it.

\begin{table*}[h]
\centering
\caption{Wilcoxon Signed-Rank Test results for all code elements between before-after versions of reusability commits.}
\begin{tabular}{|l|l|l|l|}
\hline
\textbf{Metric}                   & \textbf{p-value} & \textbf{Impact} &\textbf{Reject \(H_0\)?} \\ \hline
Percent Lack of Cohesion (LCOM)   &    0.0707     & +ve  &    False       \\ \hline
Response for Class (RFC)        &  0.2925   &  No &  False      \\ \hline
Cyclomatic Complexity (CC) & 0.3298 & +ve & False  \\ \hline
Coupling Between Objects (CBO)    &   0.2739  & +ve  &   False          \\ \hline
Weighted Method per Class (WMC)            &    0.0372      &  -ve   &    True          \\ \hline
Line of Code (LOC)               &   0.06621    & +ve     &   False             \\ \hline
Depth of Inheritance Tree (DIT)        & 0.7446 &  No  &   False           \\ \hline
Number of Children (NOC)    &   0.5292   & No  &   False     \\ \hline
 Paths (NPATH) & 0.5 &  -ve &   False\\ \hline 
 Nesting (MaxNesting)  & 0.12 &  +ve & False \\ \hline 
 Fan-in (FANIN)  &  0.97 &  -ve &  False \\ \hline 
 Fan-out (FANOUT)  &  0.94 &  -ve &  False \\ \hline 
 Lines with Comments (CLOC)  &  0.71 &  +ve &  False \\ \hline 
 Statements (STMTC)  &  0.19 &  +ve &  False \\ \hline 
 Classes (CDL)  &  0.35 &  +ve  &   False \\ \hline 
 Instance Variables (NIV)  &  0.30 &  +ve &  False \\ \hline 
 Instance Methods (NIM)  &  0.45 &  -ve &  False \\ \hline 
 Base Classes (IFANIN)  &  1.0 &  No &  False \\ \hline 
\end{tabular}
\label{tab:rq1-stat-test-results}
\end{table*}

\begin{tcolorbox}
\textbf{Summary.} When developers refactor their code for the purpose of reusability, we found that the number of methods significantly increased, but the majority of the state-of-the-art metrics did not capture any improvement, or captured non significant improvement. 
\end{tcolorbox}

\subsection{\textbf{RQ3. What triggers developers to refactor the code for the purpose of code reuse?}} \label{RQ3}
This research question aims at understanding the development contexts that trigger developer to perform refactoring activities for the purpose of code reuse. \textcolor{black}{Upon the manual inspection of the reusability refactoring commits performed by the two authors, we identify and categorize the context of the code reuse used to describe the motivation behind
the refactoring operations into nine main categories}:

\begin{itemize}
    \item  Design Patterns
    \item  Duplicate Code Removal
    \item  API Management
     \item  Features Updates
    \item  Bug Fix
    \item  Extract Reusable Component
     \item  Test Code Management
    \item  Visibility Changes
    \item  \textcolor{black}{Other refactoring operations}
\end{itemize}

\begin{figure*}[h]
\centering 
\begin{tikzpicture}
\begin{scope}[scale=0.85]
\pie[rotate = 180,pos ={0,0},text=inside, outside under=50,no number]{8.05/Design Patterns\and8.05\%,6.60/Duplicate Code Removal\and6.60\%, 3.77/API Management\and3.77\%, 12.64/Features Updates\and12.64\%,14.34/Bug Fix\and14.34\%,23.96/Extract Reusable\and23.96\%,8.11/Test Code Management\and8.11\%, 2.58/Visibility Changes\and2.58\%, 19.94/Others\and19.94\%}
\end{scope}
\end{tikzpicture}
\caption{
Distribution of code reuse context in reusability refactoring commits.}
\label{fig:codereusecontext}
\end{figure*}
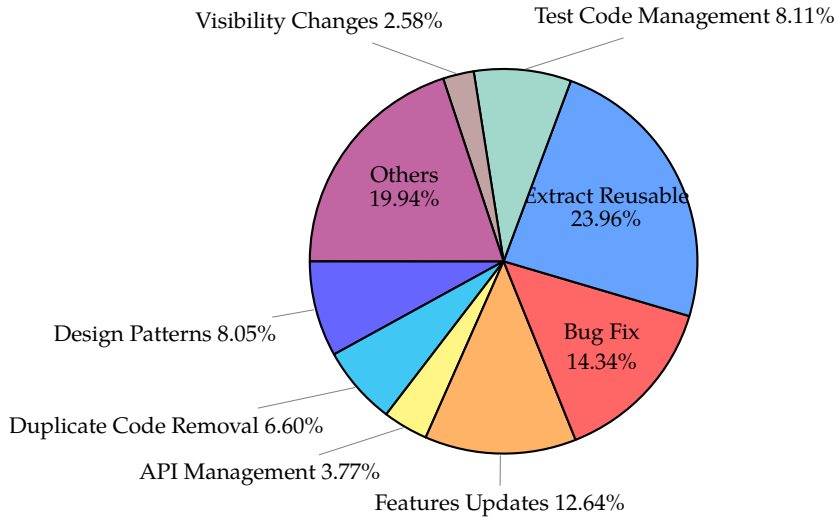

\begin{figure*}[h]
\begin{center}
\includegraphics[width=1.0\textwidth]{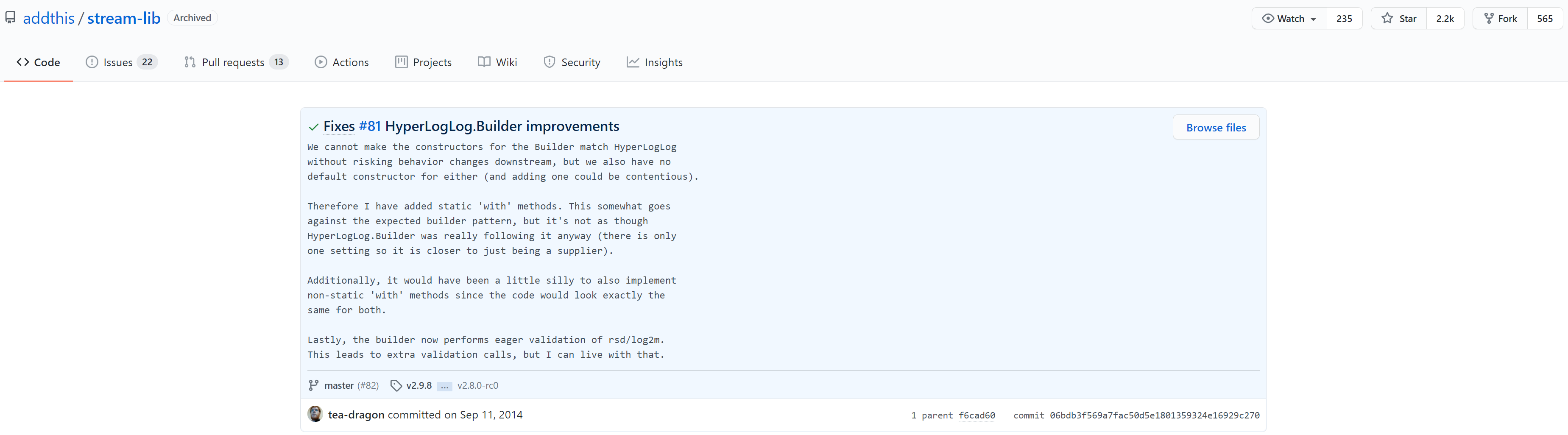}

\caption{
Commit message stating the implementation of a Builder design pattern \cite{stream-lib}.}
\label{fig:case study design pattern commit message}
\end{center}
\end{figure*}

Figure \ref{fig:codereusecontext} depicts the categorization of reusability refactoring commits. The Extract Reusable Components category had the highest number of commits with 23.96\% followed by Others and Bug Fix categories with a slight advantage to the first category, since its ratio was 19.94\%, while the Bug Fix category had a ratio of 14.34\%. We observe that a few categories had almost a uniform distribution of refactoring classes with low variability. For instance, Test Code Management, Design Patterns, and Duplicate Code Removal had  commit message distribution percentages of 8.11\%, 8.05\%, and 6.60\%, respectively. The API Management and Visibility Changes had the least popular categories. In the following section, we provide a case study for design patterns, duplicate code removal, and API management categories due to their popularity within the context of code reuse.

\subsubsection{Design Patterns.}

Design patterns are technique for achieving reuse of software architectures \cite{gamma1995elements}. Refactoring is another technique used for producing better maintainable and reusable designs. Design patterns and refactoring are powerful techniques for code reuse and are also related in the sense that design pattern can be used to guide refactoring \cite{quan2008formal}. Fowler et al. \cite{fowler2018refactoring} demonstrated such correlation by illustrating how the State Pattern is used to guide transformations of a program step by step using refactoring rules. Further, Kerievsky \cite{kerievsky2005refactoring} discussed the pattern-directed refactoring, which can be seen as big-step refactoring rules toward patterns. In the Gang-of-Four book \cite{gamma1995elements}, design patterns are classified into three categories:
Creational, Structural and Behavioral patterns. In this paper, we will take one representative pattern
from the Creational design pattern category to demonstrate the case of code reuse.

\noindent
\textbf{\textit{Case study.}} 
Builder pattern is considered one of the creational design patterns that helps construct complex objects step by step. The pattern allows to produce different types and representations of an object using the same construction code. The commit message (see Figure \ref{fig:case study design pattern commit message}) indicates improvement in the builder class by making changes in the code by using Builder pattern. The developer is trying to match the constructors of the two classes \texttt{HyperLogLog} and \texttt{Builder} without changing the behaviour of the flow of code. The developer instead adds a private constructor to try and reduce the risk of behaviour. Also, another interesting comment in this commit is:
\begin{quote}
\par 
\textit{`This somewhat goes against the expected Builder pattern, but it’s not as though HyperLogLog. Builder was really following it anyway}'
\end{quote}
This indicates that they have used a Builder design pattern although there is a possibility that it has been tweaked according to the developer’s needs. This shows the versatility of the developers and how design patterns can be changed according to one’s need.

After further investigation we found that the developer has used the Builder design pattern in the code although it has been changed to accommodate the developer’s needs. While we have a reader class \texttt{Builder} and the Builder interface \texttt{IBuilder} interface, the \texttt{Builder} class is embedded in the \texttt{HyperLogLog class}. The \texttt{IBuilder} interface consists of method calls \texttt{build()} and \texttt{sizeOf()}, in Listing \ref{IBuilder}, which are implemented in the \texttt{Builder} class. The \texttt{build()} method is of type \texttt{HyperLogLog} while \texttt{sizeOf()} is of Type \texttt{int} as shown in the Listing \ref{BuildSizeoff}.

\begin{figure*}[h]
\centering 
\includegraphics[width=1.0\textwidth]{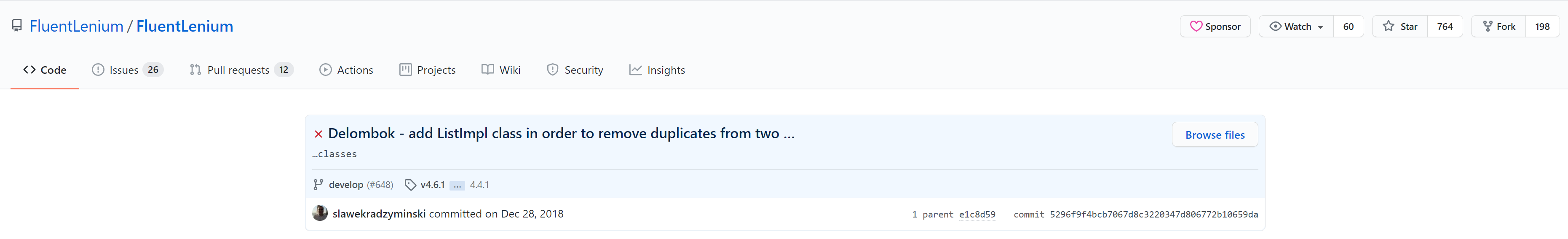}
\caption{
Commit message stating the removal of a duplicate code \cite{FluentLenium}.}
\label{fig:case study duplicate code commit message}
\end{figure*}

\begin{figure*}[htbp]
\centering
\begin{subfigure}{6cm}
\centering\includegraphics[width=6cm]{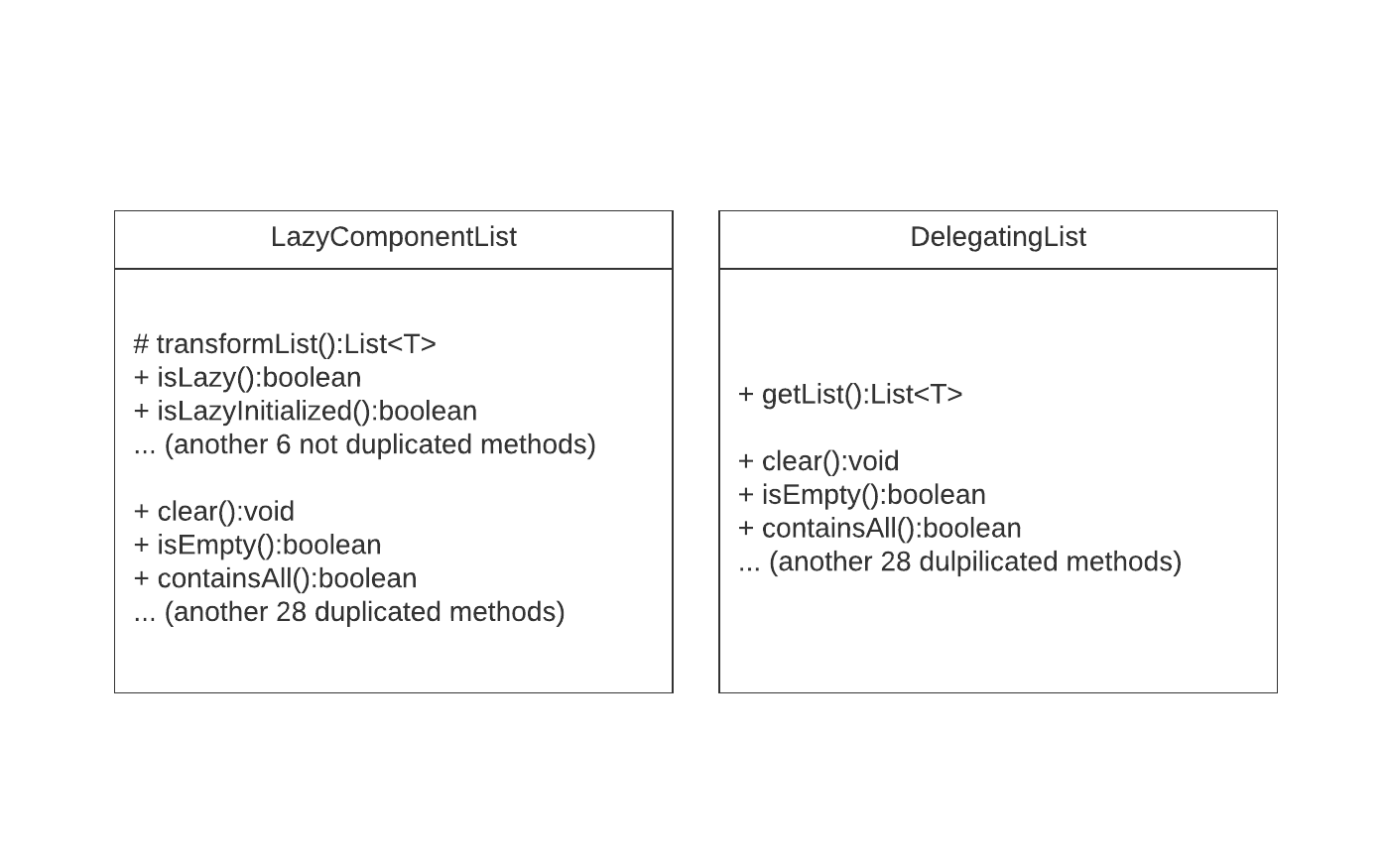}
\caption{Before Refactoring}
\label{UML:DR-v1}
\end{subfigure}%
\begin{subfigure}{6cm}
\centering\includegraphics[width=6cm]{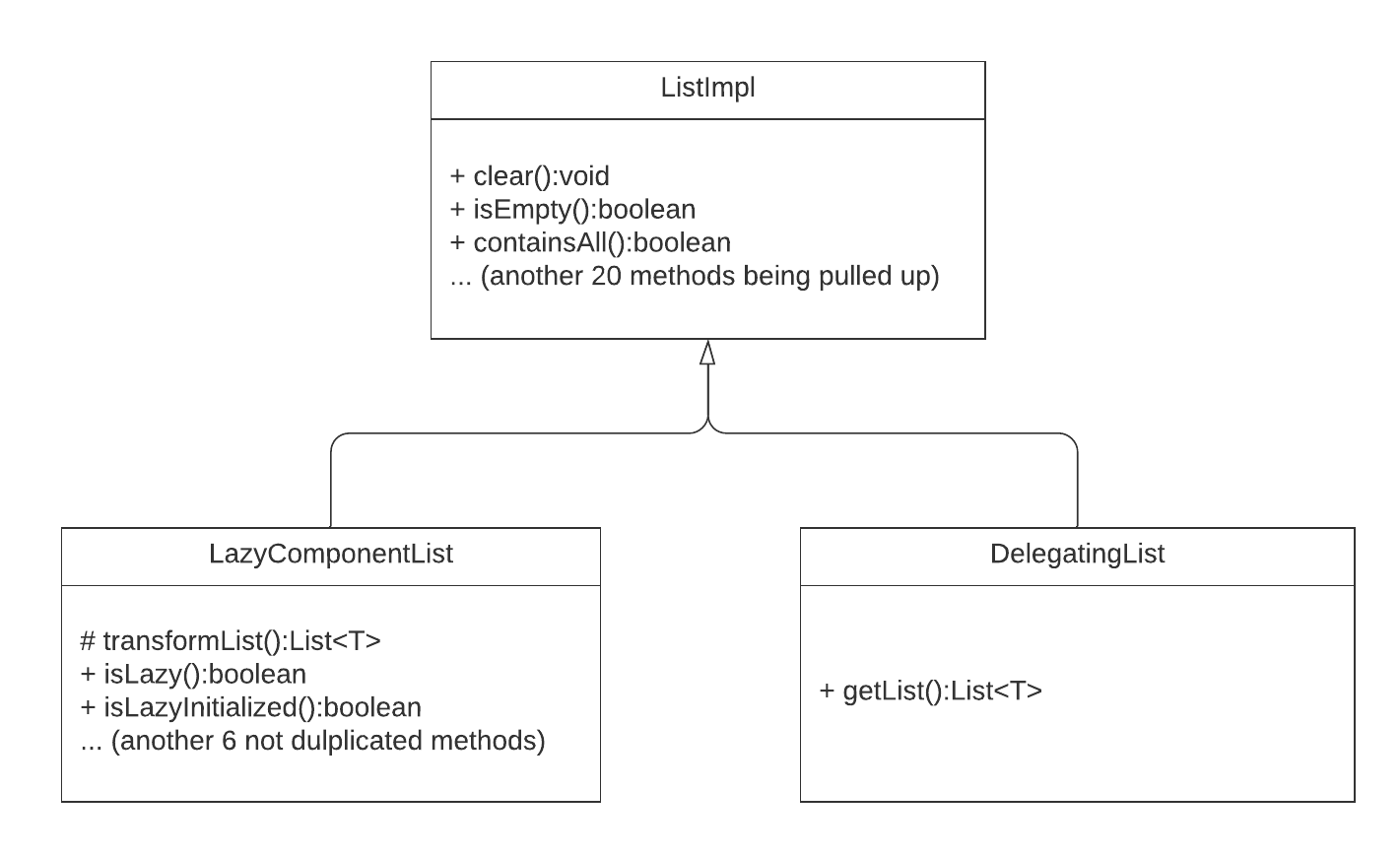}
\caption{After Refactoring}
\label{UML:DR-v2}
\end{subfigure}

\caption{
Duplicate code removal after the application of refactoring.}
\label{fig:rq3-duplicate-code-uml}
\end{figure*}

\begin{figure}[htbp]
\noindent\begin{minipage}[t]{.48\textwidth}
\begin{lstlisting}[caption={IBuilder Interface},label=IBuilder, frame=single,breaklines=true,language=diff]{Name}@@ -85,8 +85,8 @@
package com.clearspring.analytics.util;

public interface IBuilder<T> {
T build();
int sizeof();
}
    \end{lstlisting}
\end{minipage}%
\end{figure}

\begin{figure}[htbp]
\noindent\begin{minipage}[t]{.48\textwidth}
\begin{lstlisting}[caption={Implementation of build() and sizeof()},label=BuildSizeoff, frame=single,language=diff]{Name}@@ -85,8 +85,8 @@
public static class Builder implements IBuilder
<ICardinality>, Serializable {
-        private double rsd;
+        private final int log2m;
+        /**
+         * Uses the given RSD percentage to determine 
+         * how many bytes the constructed HyperLogLog 
+         * will use.
+         *
+         * @deprecated Use {@link #withRsd(double)} 
+         * instead. This builder's constructors did 
+         * not match the (already
+         * themselves ambiguous) constructors of the 
+        * HyperLogLog class, but there is no way to 
+         * make them match without
+         * risking behavior changes downstream.
+         */
+        @Deprecated
        public Builder(double rsd) {
-       this.rsd = rsd;
+        this(log2m(rsd));
+        }
+        /** This constructor is private to prevent 
+        behavior change for ambiguous usages. 
+        (Legacy support). */
+        private Builder(int log2m) {
+        validateLog2m(log2m);
+        this.log2m = log2m;
        }
        @Override
        public HyperLogLog build() {
-        return new HyperLogLog(rsd);
+        return new HyperLogLog(log2m);
        }
        @Override
        public int sizeof() {
-        int log2m = log2m(rsd);
        int k = 1 << log2m;
        return RegisterSet.getBits(k) * 4;
        }
    \end{lstlisting}
\end{minipage}%
\end{figure}

\begin{figure*}[htbp]
\noindent\begin{minipage}[t]{.48\textwidth}
\begin{lstlisting}[caption={Before refactoring},label=DPBEFORE, frame=single,breaklines=true,language=diff]{Name}@@ -85,8 +85,8 @@
    @Deprecated
    public HyperLogLog(int log2m, RegisterSet 
    registerSet) {
        if (log2m < 0 || log2m > 30) {
           throw new IllegalArgumentException
           ("log2m argument is "
             + log2m + " and is outside the range 
         [0, 30]"); }

        this.registerSet = registerSet;
        this.log2m = log2m;
        int m = 1 << this.log2m;
        alphaMM = getAlphaMM(log2m, m);
    }
    }
*
*
*
* 
*
*
*
*
*
*
*
* 
*
*
\end{lstlisting}
\end{minipage}%
\hfill
%
\begin{minipage}[t]{.48\textwidth}
\begin{lstlisting}[caption={After refactoring},label=DPAFTER,frame=single, breaklines=true,language=diff]{Name}@@ -85,8 +85,8 @@
+   private static void validateLog2m(int log2m) {
+        if (log2m < 0 || log2m > 30) {
+            throw new IllegalArgumentException
+             ("log2m argument is " + log2m + " and is 
+             outside the range [0, 30]");
+        }
+    }
+
    /**
     * Create a new HyperLogLog instance.  The log2m parameter defines the accuracy of
     * the counter.  The larger the log2m the better 
      the accuracy.
@@ -117,10 +124,7 @@ public HyperLogLog(int log2m) {
     */
    @Deprecated
    public HyperLogLog(int log2m, RegisterSet 
    registerSet) {
-        if (log2m < 0 || log2m > 30) {
-            throw new IllegalArgumentException
-            ("log2m argument is "
-             + log2m + " and is outside the range 
-         [0, 30]"); }
+        validateLog2m(log2m);
        this.registerSet = registerSet;
        this.log2m = log2m;
        int m = 1 << this.log2m;
        alphaMM = getAlphaMM(log2m, m);
    }

\end{lstlisting}
\end{minipage}%

\end{figure*}

\begin{figure}[htbp]
\noindent\begin{minipage}[t]{.48\textwidth}
\begin{lstlisting}[caption={validateLog2m() is called in Builder() constructor},label=validateLog2m, frame=single,breaklines=true,language=diff]{Name}@@ -85,8 +85,8 @@
public static class Builder implements IBuilder
<ICardinality>, Serializable {
-        private double rsd;
+        private final int log2m;
+        /**
+         * Uses the given RSD percentage to determine 
+         * how many bytes the constructed HyperLogLog 
+         * will use.
+         *
+         * @deprecated Use {@link #withRsd(double)} 
+         * instead. This builder's constructors did 
+         * not match the (already
+         * themselves ambiguous) constructors of the 
+        * HyperLogLog class, but there is no way to 
+         * make them match without
+         * risking behavior changes downstream.
+         */
+        @Deprecated
        public Builder(double rsd) {
-       this.rsd = rsd;
+        this(log2m(rsd));
+        }
+        /** This constructor is private to prevent 
+        behavior change for ambiguous usages. 
+        (Legacy support). */
+        private Builder(int log2m) {
+        validateLog2m(log2m);
+        this.log2m = log2m;
        }

    \end{lstlisting}
\end{minipage}%
\end{figure}

The refactoring operations performed in this commit are \textit{Extract Method} and \textit{Rename Method}. \textit{Extract Method} is performed on the function public \texttt{HyperLogLog()} from which the code snippet, as shown in Listings \ref{DPBEFORE} and \ref{DPAFTER}, respectively, is extracted to a new function \texttt{private static void validateLog2m()}. This function is then called in the \texttt{public HyperLogLog()} and also in the private \texttt{Builder()} constructor as shown in in Listing \ref{validateLog2m}. This refactoring operation can be considered as reusability, instead of writing the code snippet twice, it can be simplified and made more efficient by calling a function. The second refactoring operation that has been performed is renaming an attribute, indicated in Listing \ref{validateLog2m}. The variable \texttt{private double rsd} is renamed to \texttt{private final int log2m} to keep the naming convention consistent with the rest of the code.

\subsubsection{Duplicate Code Removal.}

Reuse of code fragments by
copying and pasting from one location to another is very
common approach during software development. Code duplication is generally discouraged as it might make the software maintenance difficult \cite{hotta2010duplicate,roy2014vision,mondal2011empirical}. For instance, if the existing code fragment is copied, bugs need to be fixed at multiple places, which makes the task inefficient and error-prone.


\begin{figure*}[htbp]
\noindent\begin{minipage}[t]{.48\textwidth}
\begin{lstlisting}[caption={DelegatingList Class},label=DelegatingList, frame=single,breaklines=true,language=diff]{Name}@@ -85,8 +85,8 @@
public void clear() {
    getList().clear();
    }
    public void forEach(Consumer<? super T> action) {
    getList().forEach(action);
    }
    public <T> T[] toArray(IntFunction<T[]> generator) 
    {
    return getList().toArray(generator);
    }
    public boolean isEmpty() {
    return getList().isEmpty();
    }
    public boolean removeIf(Predicate<?super T>filter)
    {
    return getList().removeIf(filter);
    }
    public Spliterator<T> spliterator() {
    return getList().spliterator();
    }
    public T set(int index, T element) {
    return getList().set(index, element);
    }
    public boolean containsAll(Collection<?> c) {
    return getList().containsAll(c);
    }
\end{lstlisting}
\end{minipage}%
\hfill
%
\begin{minipage}[t]{.48\textwidth}
\begin{lstlisting}[caption={LazyComponentList Class},label=LazyComponentList,frame=single, breaklines=true, language=diff]{Name}@@ -85,8 +85,8 @@
  public void clear() {
    getList().clear();
    }
    public void forEach(Consumer<?super T> action) {
    getList().forEach(action);
    }
    public <T> T[] toArray(IntFunction<T[]> generator) 
    {
    return getList().toArray(generator);
    }
    public boolean isEmpty() {
    return getList().isEmpty();
    }
    public boolean removeIf(Predicate<?super T>filter) 
    {
    return getList().removeIf(filter);
    }
    public Spliterator<T> spliterator() {
    return getList().spliterator();
    }
    public T set(int index, T element) {
    return getList().set(index, element);
    }
    public boolean containsAll(Collection<?> c) {
    return getList().containsAll(c);
    }
\end{lstlisting}
\end{minipage}%

\end{figure*}

\begin{figure*}[htbp]
\noindent\begin{minipage}[t]{.48\textwidth}
\begin{lstlisting}[caption={LazyComponentList Class (before removing duplicate)},label=LazyComponentListBEFORE, frame=single,breaklines=true,language=diff]{Name}@@ -85,8 +85,8 @@
- public class LazyComponentList<T> implements 
- List<T>, WrapsElements, LazyComponents<T> {
    private final ComponentInstantiator instantiator;
    private final Class<T> componentClass;

@@ -120,127 +110,4 @@ public String toString() {
    return isLazyInitialized() ? getList().toString() 
    : elements.toString();
    }
-    public void clear() {
-    getList().clear();
-    }
-    public void forEach(Consumer<? super T> action) {
-    getList().forEach(action);
-    }
-    public <T> T[] toArray(IntFunction<T[]> generator) 
-    {
-    return getList().toArray(generator);
-    }
-    public boolean isEmpty() {
-    return getList().isEmpty();
-    }
-    public boolean removeIf(Predicate<?super T>filter) 
-    {
-    return getList().removeIf(filter);
-    }
-    public Spliterator<T> spliterator() {
-    return getList().spliterator();
-    }
-    public T set(int index, T element) {
-    return getList().set(index, element);
-    }
-    public boolean containsAll(Collection<?> c) {
-   return getList().containsAll(c);
-    }
\end{lstlisting}
\end{minipage}%
\hfill
%
\begin{minipage}[t]{.48\textwidth}
\begin{lstlisting}[caption={LazyComponentList Class (after removing duplicate)},label=LazyComponentListAFTER,frame=single,breaklines=true, language=diff]{Name}@@ -85,8 +85,8 @@
+ public class LazyComponentList<T> extends 
+ ListImpl<T> implements List<T>, WrapsElements, 
+ LazyComponents<T> 
{
    private final ComponentInstantiator instantiator;
    private final Class<T> componentClass;

@@ -120,127 +110,4 @@ public String toString() {
    return isLazyInitialized() ? getList().toString() 
    : elements.toString();
    }
    }
*
*
*
* 
*
*
*
*
*
*
*
* 
*
*
*
*
*
*
* 
*
*
*
*
*    
*
*
\end{lstlisting}
\end{minipage}%

\end{figure*}

\begin{figure*}[htbp]
\noindent\begin{minipage}[t]{.48\textwidth}
\begin{lstlisting}[caption={DelegatingList Class (before removing duplicate)},label=DelegatingListBEFORE, frame=single,language=diff]{Name}@@ -85,8 +85,8 @@
- public class DelegatingList<T> implements List<T> {
    protected final List<T> list;
    /**
@@ -34,127 +24,4 @@ public DelegatingList(List<T> list) 
    {
    return list;
    }
-   public void clear() {
-    getList().clear();
-    }
-    public void forEach(Consumer<? super T> action) {
-    getList().forEach(action);
-    }
-    public <T> T[] toArray(IntFunction<T[]> generator) 
-    {
-    return getList().toArray(generator);
-    }
-    public boolean isEmpty() {
-    return getList().isEmpty();
-   }
-    public boolean removeIf(Predicate<?super T>filter) 
-    {
-    return getList().removeIf(filter);
-    }
-    public Spliterator<T> spliterator() {
-    return getList().spliterator();
-    }
-    public T set(int index, T element) {
-    return getList().set(index, element);
-    }
-    public boolean containsAll(Collection<?> c) {
-    return getList().containsAll(c);
-    }
\end{lstlisting}
\end{minipage}%
\hfill
%
\begin{minipage}[t]{.48\textwidth}
\begin{lstlisting}[caption={DelegatingList Class (after removing duplicate)},label=DelegatingListAFTER,frame=single, language=diff]{Name}@@ -85,8 +85,8 @@
+ public class DelegatingList<T> extends ListImpl<T> 
+ implements List<T> {
    protected final List<T> list;

    /**
@@ -34,127 +24,4 @@ public DelegatingList(List<T> list) 
    {
        return list;
    }
*
*
*
*
*
*
*
*
*
*
*
*
*
*
*
*
*
*
*
*
*
*
*
*



\end{lstlisting}
\end{minipage}%

\end{figure*}

\begin{figure}[htbp]
\noindent\begin{minipage}[t]{.48\textwidth}
\begin{lstlisting}[caption={ListImpl Class},label=ListImpl, frame=single,language=diff]{Name}@@ -85,8 +85,8 @@
+ public abstract class ListImpl<T> implements List<T> 
+   {
+    public ListImpl() {
+    super();
+   }
+    public abstract List<T> getList();
+    public void clear() {
+    getList().clear();
+    }
+    public boolean isEmpty() {
+    return getList().isEmpty();
+    }
+    public T set(int index, T element) {
+    return getList().set(index, element);
+    }
+    public boolean containsAll(Collection<?> c) {
+    return getList().containsAll(c);
+    }
+    public List<T> subList(int fromIndex, int toIndex) 
+    {
+    return getList().subList(fromIndex, toIndex);
+    }
+    public boolean add(T e) {
+    return getList().add(e);
+    }
+    public boolean remove(Object o) {
+    return getList().remove(o);
+    }
    \end{lstlisting}
\end{minipage}%
\end{figure}

\noindent
\textbf{\textit{Case study.}} In this case study we are investigating the refactoring operations performed by the developers which claim to remove duplicate codes in their git commits. Extending our research in reusability refactoring commits, we venture into the domain of code smells and try to find instances where refactoring operations could aid in improving not only the quality of code but also remove code smells to some extent. 
The commit message in Figure \ref{fig:case study duplicate code commit message},  indicates refactoring operation being performed to remove duplicates from code. \textit{Extract Superclass} is a common refactoring operation for removing the clones while two classes share some common methods. For example, the detected refactorings are \textit{Extract Superclass} operation followed with several \textit{Pull Up Method} operations.

\begin{figure*}[h]
\centering 
\includegraphics[width=1.0\textwidth]{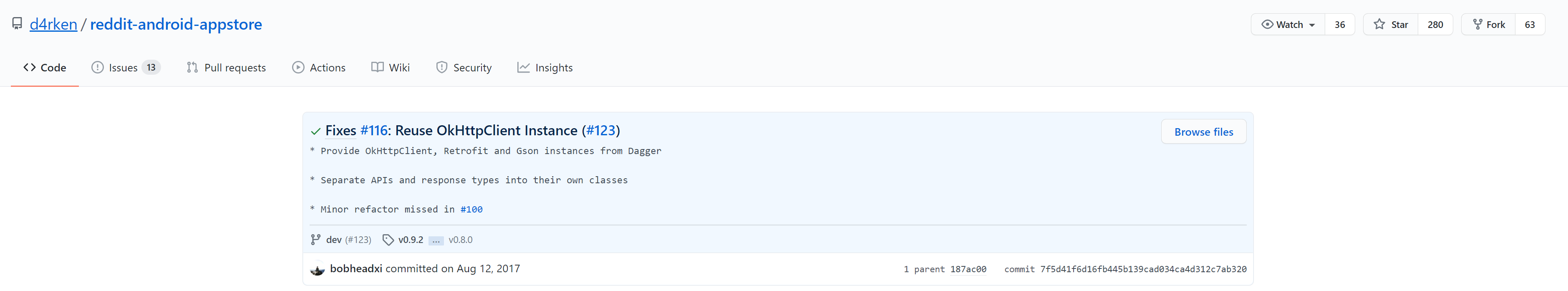}
\caption{
Commit message stating the management of the API \cite{d4rken}.}
\label{fig:case study API management}
\end{figure*}

By looking into the actual code fragments, Listings \ref{DelegatingList} and \ref{LazyComponentList} show that the \texttt{DelegatingList} class and \texttt{LazyComponentList} class have duplicated methods. Note that there are totally 31 duplicated methods before the refactoring, we only show parts of them as examples for a better visualization. To fix this, the developer introduces a new class \texttt{ListImpl}, pulls up those duplicated methods to \texttt{ListImlp}, then makes both \texttt{DelegatingList} and  \texttt{LazyComponentList} extends \texttt{ListImpl}. Listings \ref{LazyComponentListBEFORE}, \ref{LazyComponentListAFTER} and \ref{DelegatingListBEFORE} show part of the refactored codes from Github. We can see that in Listings \ref{LazyComponentListBEFORE} and \ref{LazyComponentListAFTER} the duplicated methods such \texttt{clear}, \texttt{isEmpty} and \texttt{containsAll} are being removed, and the \texttt{DelegatingList} class and \texttt{LazyComponentList} class now extends the newly introduced \texttt{ListImpl} class. Listing \ref{DelegatingListBEFORE} shows that the methods being removed in Listings \ref{LazyComponentListBEFORE} and \ref{LazyComponentListAFTER} are now pulled up to the superclass \texttt{ListImpl}. Listings \ref{DelegatingListAFTER} and \ref{ListImpl} show the UML diagrams before the commit and after the commit. Note that there are 31 duplicated methods before the refactoring, and only 23 of them are being pulled up. That is because there are 8 unnecessary methods overloading being permanently removed after refactoring.

The UML diagrams (see Figures \ref{UML:DR-v1} and \ref{UML:DR-v2}) clearly show that the clones have been removed by the \textit{Extract Superclass} operation and the \textit{Pull Up Method} operations. This case study gives us an example that the refactorings removing the code duplication such as \textit{Extract Superclass} and \textit{Pull Up Method} could be helpful for improving the reusability. With those duplicated methods being pulled to a superclass, if a new type of List is introduced, it can reuse those methods since they are common list operations. 


\subsubsection{API Management.}
Dig et al. \cite{dig2005role} studied the role of refactoring in API evolution with the goal of reducing the burden of reuse on maintenance. This requires either minimizing the amount of change or reducing the cost of adapting to change. For instance, when a class library that is reused in many systems
is refactored, the systems that reuse it must change.

\textbf{\textit{Case study.}} Figure \ref{fig:case study API management} provides an example of API management code reuse context. Refactoring Miner detects 13 refactoring operations. Although this is a large number of refactoring operations, only 3 operations (i.e., \textit{Move Class}) are directly related to API management. The commit message shows that the intention of the refactoring is to separate APIs into their own classes. Before the refactoring, the APIs are in their corresponding Repository classes which is a bad practice since the Repository classes are handling extra functionalities that should not belong to them. Separating APIs into their own classes will reduce the coupling in the code and allow the developer to efficiently reuse the APIs and their functionalities. By looking at the refactoring in details, we can see that the \texttt{GithubApi}, and \texttt{TokenApi} 
are being refactored with the move class operations. These API are being extracted and moved from their original Repository classes to their own classes to improve code reusability. Next, we will explain each refactored APIs to help better understand the developer’s intention.

Listings \ref{BefacorRefact-LiveGithubRepository.java}, \ref{BeforeRefact-TokenRepository.java}, 
show that before the refactoring \texttt{GithubApi}, \texttt{TokenApi}, 
interface are part of \texttt{LiveWikiRepository.java}\footnote{Full file path: LiveGithubRepository.java/TokenRepository.java/LiveWikiRepository.java} and is highly coupled with it, which is bad for code reuse. Listings \ref{AfterRefact-GithubApi.java}. \ref{after refactoring - In TokenApi.java}, 
show after the refactoring, for better reuse of the \texttt{GithubApi}, \texttt{TokenApi}, 
interface, it is being extracted and moved to \texttt{WikiApi.java}\footnote{Ful file path: GithubApi.java/TokenApi.java/WikiApi.java}. In this case, like Listings\ref{Github API (before)} and \ref{Github API (after)} show, when the other classes need \texttt{GithubApi}, \texttt{TokenApi}, 
they can just import and use it instead of using the whole \texttt{LiveGithubRepository}, \texttt{TokenRepository}, 
that are loaded with redundant functionalities.

\subsubsection{\textcolor{black}{Feature Updates.}}

\textcolor{black}{Code reuse can be used for the purpose of adding a new function or update an existing one.}

\begin{figure*}[h]        
\centering 
\includegraphics[width=1.0\textwidth]{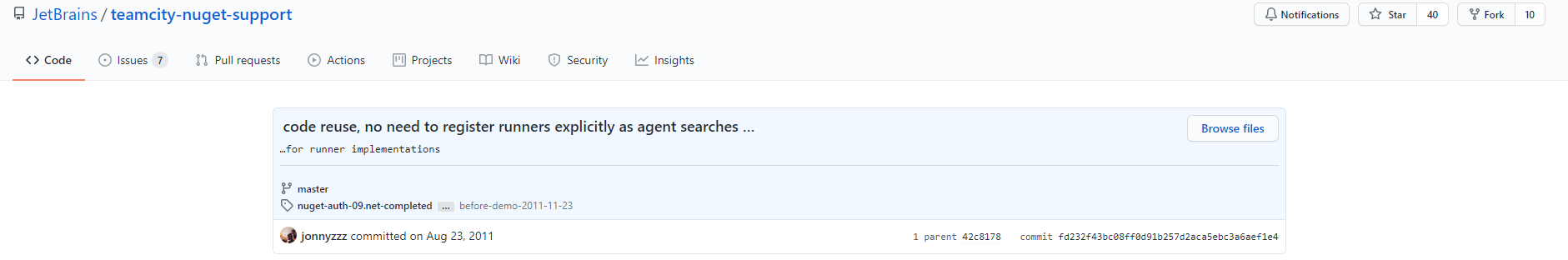}
\caption{
\textcolor{black}{Commit message stating the update of a feature \cite{JetBrains}.}}
\label{fig:case study feature update}
\end{figure*}

\textcolor{black}{\textbf{\textit{Case study.}} Figure \ref{fig:case study feature update} shows the commit message and the intentions of the developer which clearly state the purpose of this refactoring operation, that is, code reusability. The developer explains the need to no longer register runners explicitly as agent searches for runner implementations, which can mean that the developer intended to remove these runner implementations from the base classes and store them separately which can be called implicitly as and when needed.}

\textcolor{black}{In Listings \ref{PackagesInstallerRunner}, \ref{PackagesPublishRunner}, and \ref{PackRunner}, we can see that all these classes, namely: \texttt{PackagesInstallerRunner},  \texttt{PackagesPublishRunner} and \texttt{PackRunner} respectively implement interfaces \texttt{AgentBuildRunner} and \texttt{AgentBuildRunnerInfo} provided by jetbrains, that explicitly running function \texttt{getRunnerInfo()} provided by \texttt{AgentBuildRunnerInfo()} and \texttt{canRun()}.}

\textcolor{black}{Apart from the two functions the constructors in each class in Listings \ref{PackagesInstallerRunner}, \ref{PackagesPublishRunner}, and \ref{PackRunner}, repeat variables \texttt{myNuGetActionFactory} and \texttt{myNuGetActionFactory}} \\
\textcolor{black}{which can be considered as repeating code or duplicate code. This can affect the time consumed by the program overall at runtime as the program runs all these lines of code even when not required. Thus, the developer extracts these repeating components from the classes in Listings \ref{PackagesInstallerRunner}, \ref{PackagesPublishRunner}, and \ref{PackRunner} and moves them to a superclass \texttt{NuGetRunnerBase} in Listing \ref{NuGetRunnerBase}. This superclass is then extended by each class and this superclass implements the interfaces \texttt{AgentBuildRunner} and \texttt{AgentBuildRunnerInfo} provided by jetbrains. This is considered an example of feature update, as the features used from the jetbrains interface classes and the common functionalities among the classes: \texttt{PackagesInstallerRunner}, \texttt{PackagesPublishRunner} and  \texttt{PackRunner}, are now being used by each class implicitly via the  \texttt{NuGetRunnerBase}, thus promoting reusability as well.}

\subsubsection{\textcolor{black}{Bug Fix.}}

\textcolor{black}{Due to the habit of copy-and-paste in programming, similar code fragments may contain the same bug which may be
neglected to fix during software maintenance. Developers can reuse existing code and interleave it with fixing the bug.}

\begin{figure*}[h]
\centering 
\includegraphics[width=1.0\textwidth]{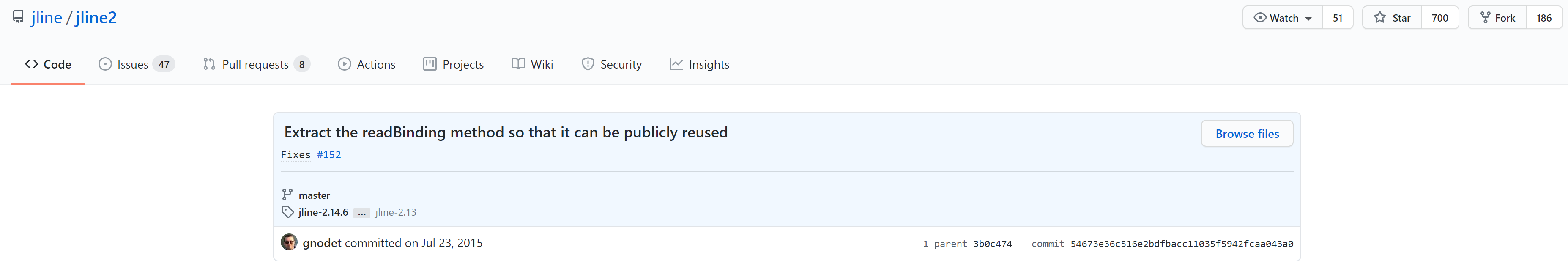}
\caption{
\textcolor{black}{Commit message stating the fixing of a bug \cite{jline}.}}
\label{fig:case study bugfix}
\end{figure*}

\textcolor{black}{\textbf{\textit{Case study.}} This case study provides an example of Bug Fix context. The commit (see Figure \ref{fig:case study bugfix}) is in project JLine which is a Java library for handling console input. A user reported a bug in the code that \textit{the arrow keys are all reported as ASCII code 27 with the \texttt{readCharacter} method}. The problem behind this is there is no real ASCII code for the arrow keys, and they are usually represented by two characters rather than one character. One of the possible solutions is to allow the user to define their own key bindings in KeyMap, then while reading an input stream, wait until a valid binding is found in KeyMap. Currently, method \texttt{readLine} contains the statements related to processing the pre-defined key bindings, and there is no access point for the user to use their own key bindings. To decouple the functionality of processing key bindings from a large method and improve the code reusability, the developer decides to extract those statements from the \texttt{readLine} method.}

\textcolor{black}{As Listings \ref{BFBEFORE} and \ref{BFAFTER} show, the developer extracted the part processing the key bindings in method \texttt{readLine} to a new method \texttt{readBinding} which can be publicly reused. The \texttt{readBinding} method will take a KeyMap as an argument, which could be defined by the user. It will block until there is a matching binding found or reach the end of line, so it can support reading the two characters arrow keys.}

\subsubsection{\textcolor{black}{Extract Reusable Component.}}

\textcolor{black}{Extracting code elements can be performed with reusability in mind and there is a high possibility that used components of the code were extracted.}

\begin{figure*}[h]
\centering 
\includegraphics[width=1.0\textwidth]{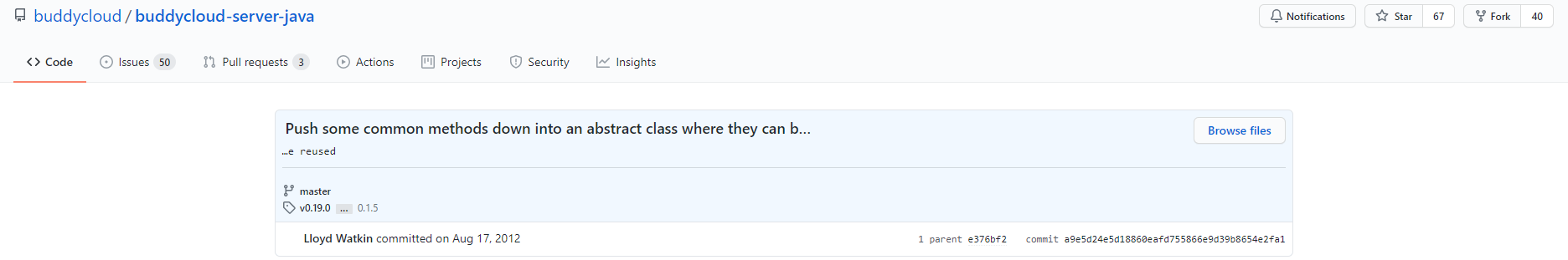}
\caption{
\textcolor{black}{Commit message stating the extraction of reusable component \cite{buddycloud}.}}
\label{fig:case study extract reusable component}
\end{figure*}

\textcolor{black}{\textbf{\textit{Case study.}} The developer mentions pushing some common methods down into an abstract class in Figure \ref{fig:case study extract reusable component}, which can be considered as an extract refactoring operation and thus help improve reusability as the code can now be used by other classes as well and the implementations can be easily changed in a common file unless overridden. This saves the developer time and makes programming very easy and structured.}

\textcolor{black}{As shown in Listing \ref{PubSubElementProcessorAbstract}, the methods in class PubSubElementProcessorAbstract have been extracted from class NodeCreate as shown in Listings \ref{PubSubElementProcessorAbstract} and \ref{NodeCreate}.  Upon further investigation, we checked if there were other classes that extended the  \texttt{PubSubElementProcessorAbstract} class and we found multiple other classes that extend this class. Also, we searched for any functions used from \texttt{PubSubElementProcessorAbstract} class in any of the classes that extend it and found a few of classes,  which show the use of functions \texttt{getNodeConfigurationHelper()} and \texttt{setOutQueue()}. 
}

\subsubsection{\textcolor{black}{Test Code Management.}}

\textcolor{black}{Code reuse helps in reducing testing efforts. Reusing the code fragments, which have already been unit tested, reduces and saves testing effort, both in terms of avoiding the need to write additional tests and also by not having to run those additional tests each time the full test suite is run.}

\begin{figure*}[h]
\centering 
\includegraphics[width=1.0\textwidth]{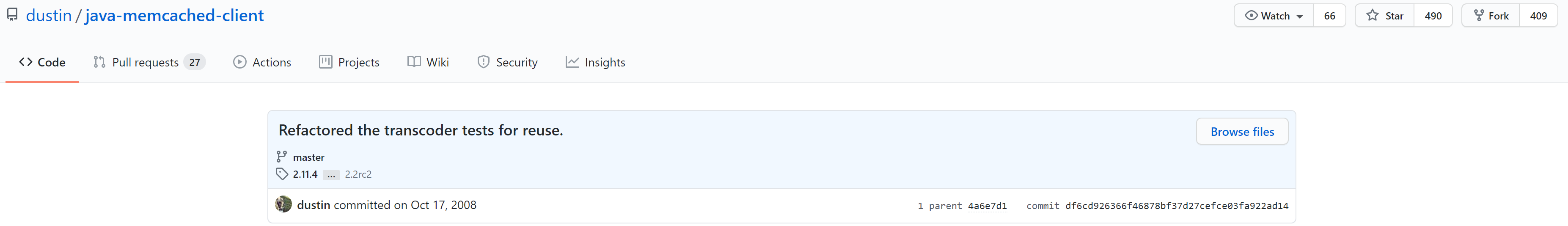}
\caption{
\textcolor{black}{Commit message stating management of the test code \cite{dustin}.}}
\label{fig:case study test}
\end{figure*}

\textcolor{black}{\textbf{\textit{Case study.}} Figure \ref{fig:case study test} provides an example of Test Code Management context in which  developer refactored the \texttt{transcoder} tests for reuse. There are 39 detected refactoring operations including 1 \textit{Extract Superclass} and 38 \textit{Pull Up Method}, where 3 classes are involved - \texttt{WhalinTranscoderTest}, \texttt{SerializingTranscoderTest} and \texttt{BaseTranscoderCase}. The \texttt{WhalinTranscoderTest} and \texttt{SerializingTranscoderTest} share 18 duplicated methods. To achieve the goal stated by developer in the commit message, the developer pulled up 18 duplicated methods, and extracted 2 methods(\texttt{testShort}, \texttt{testCharacter}) that are not duplicated but helpful for future reuse, to a newly introduced class \texttt{BaseTranscoderCase}. In the future, the classes that need to be tested could reuse the test cases in \texttt{BaseTranscoderCase} to perform some basic tests such as type check and boundary check.}

\textcolor{black}{For example, Listings \ref{Test1 (before)} and \ref{Test2 (before)} illustrate the  refactoring of 4 duplicated methods - \texttt{testLong}, \texttt{testInt}, \texttt{testChar}, \texttt{testBoolean} shared by both \texttt{SerializingTranscoderTest} and \texttt{WhalinTranscoderTest}, and there are 1 method \texttt{testShort} that is only in\texttt{ WhalinTranscoderTest}. Listing \ref{Test (after)} shows after the refactoring, all of these 5 methods are being pulled up and extracted to the new class \texttt{BaseTranscoderCase} to improve reusability. }
\begin{figure}[H]
\noindent\begin{minipage}[t]{.48\textwidth}
\begin{lstlisting}[caption={Before refactoring - LiveGithubRepository.java},label=BefacorRefact-LiveGithubRepository.java, frame=single,breaklines = true,language=diff]{Name}@@ -85,8 +85,8 @@
public class LiveGithubRepository implements 
GithubRepository {

-    interface GithubApi {
-        @GET("repos/d4rken/reddit-android-appstore
-        /releases/latest")
-        Observable<Release> getLatestRelease();
+    private Observable<GithubApi.Release> 
+    latestReleaseCache;
+    private Observable<List<GithubApi.Contributor>> 
+    latestContributorsCache;

-        @GET("repos/d4rken/reddit-android
-        appstore/contributors")
-        Observable<List<Contributor>> 
-        getContributors();
+    public LiveGithubRepository(GithubApi githubApi) {
+        this.githubApi = githubApi;
    }

    @Override
-    public Observable<Release> getLatestRelease() {
+    public Observable<GithubApi.Release> 
+    getLatestRelease() {
        if (latestReleaseCache == null) 
        latestReleaseCache
        = githubApi.getLatestRelease().cache();
        return latestReleaseCache;
    }

    @Override
-    public Observable<List<Contributor>> 
-    getContributors() {
+   public Observable<List<GithubApi.Contributor>> 
+   getContributors() {
        if (latestContributorsCache == null) 
        latestContributorsCache = githubApi
        .getContributors().cache();
        return latestContributorsCache;
    }
    \end{lstlisting}
\end{minipage}%
\end{figure}

\begin{figure}[h]
\noindent\begin{minipage}[t]{.48\textwidth}
\begin{lstlisting}[caption={After refactoring - GithubApi.java},label=AfterRefact-GithubApi.java, frame=single,breaklines=true,language=diff]{Name}@@ -85,8 +85,8 @@
+ package subreddit.android.appstore.backend.github;

+ import com.google.gson.annotations.SerializedName;

+ import java.util.Date;
+ import java.util.List;

+ import io.reactivex.Observable;
+ import retrofit2.http.GET;

+ public interface GithubApi {
+    String BASEURL = "https://api.github.com/";

+    @GET("repos/d4rken/reddit-android-appstore/
+    releases/latest")
+    Observable<Release> getLatestRelease();

+    @GET("repos/d4rken/reddit-android-appstore/
+    contributors")
+    Observable<List<Contributor>> getContributors();

+    class Release {
+        @SerializedName("url") public String 
+        releaseUrl;
+        @SerializedName("tag_name") public String 
+        tagName;
+        @SerializedName("name") public String 
+        releaseName;
+        @SerializedName("body") public String 
+        releaseDescription;
+        public boolean prerelease;
+        @SerializedName("published_at") public Date 
+        publishDate;
+        public List<Assets> assets;

+       public static class Assets {
+            @SerializedName("browser_download_url") 
+            public String downloadUrl;
+            public long size;
+        }
+    }

+    class Contributor {
+    @SerializedName("login") public String username;
+    }
+}
    \end{lstlisting}
\end{minipage}%
\end{figure}

\begin{figure}[htbp]
\noindent\begin{minipage}[t]{.48\textwidth}
\begin{lstlisting}[caption={Before refactoring - TokenRepository.java},label=BeforeRefact-TokenRepository.java, frame=single,breaklines=true,language=diff]{Name}@@ -85,8 +85,8 @@
-  interface TokenApi {
-        @FormUrlEncoded
-        @POST("api/v1/access_token")
-        Observable<Token> getUserlessAuthToken(
-           @Header("Authorization") String 
-           authentication,
-           @Field("device_id") String deviceId,
-           @Field("grant_type") String grant_type,
-           @Field("scope") String scope
-        );
-    }

    \end{lstlisting}
\end{minipage}%
\end{figure}

\begin{figure}[htbp]
\noindent\begin{minipage}[t]{.48\textwidth}
\begin{lstlisting}[caption={After refactoring - TokenApi.java},label=after refactoring - In TokenApi.java, frame=single,breaklines=true,language=diff]{Name}@@ -85,8 +85,8 @@
+ package subreddit.android.appstore.backend.reddit;


+ import io.reactivex.Observable;
+ import retrofit2.http.Field;
+ import retrofit2.http.FormUrlEncoded;
+ import retrofit2.http.Header;
+ import retrofit2.http.POST;

+ public interface TokenApi {
+    String BASEURL = "https://www.reddit.com/";

+    @FormUrlEncoded
+    @POST("api/v1/access_token")
+    Observable<TokenApi.Token> getUserlessAuthToken
+    (@Header("Authorization") String authentication,
+    @Field("device_id") String deviceId,
+    @Field("grant_type") String grant_type,
+    @Field("scope") String scope);

+    class Token {
+        String access_token;
+        String token_type;
+        long expires_in;
+        String scope;
+        final long issuedTime = 
+        System.currentTimeMillis();

+        public boolean isExpired() {
+           return System.currentTimeMillis() 
+           > issuedTime + expires_in * 1000;
+        }

+        public String getAuthorizationString() {
+            return token_type + " " + access_token;
+        }
+    }
+ }
    \end{lstlisting}
\end{minipage}%
\end{figure}

\begin{figure}[htbp]
\noindent\begin{minipage}[t]{.48\textwidth}
\begin{lstlisting}[caption={Github API (before)},label=Github API (before), frame=single,breaklines=true,language=diff]{Name}@@ -85,8 +85,8 @@
import android.support.annotation.Nullable;

- import subreddit.android.appstore.backend.github.
- GithubRepository;

import subreddit.android.appstore.util.mvp.BasePresenter;
import subreddit.android.appstore.util.mvp.BaseView;

@@ -14,22 +14,22 @@

        void selectFilter(CategoryFilter filter);

-        void showUpdateSnackbar(@Nullable 
-        GithubRepository.Release release);


        void showUpdateErrorToast();

-        void enableUpdateAvailableText(@Nullable 
-        GithubRepository.Release release);


        void showDownload(String url);

-        void showChangelog(GithubRepository.Release 
-        release);

    }

    interface Presenter extends BasePresenter<View> {
        void notifySelectedFilter(CategoryFilter 
        categoryFilter);

-   void downloadUpdate(GithubRepository.Release 
-   release);


-   void buildChangelog(GithubRepository.Release 
-   release);

    }
}

\end{lstlisting}
\end{minipage}%
\hfill
%
\begin{minipage}[t]{.48\textwidth}
\begin{lstlisting}[caption={Github API (after)},label=Github API (after),frame=single, breaklines=true,language=diff]{Name}@@ -85,8 +85,8 @@
import android.support.annotation.Nullable;

+ import subreddit.android.appstore.backend.github.
+ GithubApi;
import subreddit.android.appstore.util.mvp.BasePresenter;
import subreddit.android.appstore.util.mvp.BaseView;

@@ -14,22 +14,22 @@

        void selectFilter(CategoryFilter filter);

+        void showUpdateSnackbar(@Nullable GithubApi
+        .Release release);

        void showUpdateErrorToast();


+        void enableUpdateAvailableText(@Nullable 
+        GithubApi.Release release);

        void showDownload(String url);

+        void showChangelog(GithubApi.Release release);
    }

    interface Presenter extends BasePresenter<View> {
        void notifySelectedFilter(CategoryFilter 
        categoryFilter);

+   void downloadUpdate(GithubApi.Release release);


+   void buildChangelog(GithubApi.Release release);
    }
}

\end{lstlisting}
\end{minipage}%

\end{figure}

\begin{figure*}[htbp]
\noindent\begin{minipage}[t]{.48\textwidth}
\begin{lstlisting}[caption={\textcolor{black}{PackagesInstallerRunner Class}},label=PackagesInstallerRunner, frame=single,breaklines=true,language=diff]{Name}@@ -85,8 +85,8 @@
- public class PackagesInstallerRunner implements 
- AgentBuildRunner, AgentBuildRunnerInfo {
-   private static final Logger LOG = Logger.
-   getInstance
-  (PackagesInstallerRunner.class.getName());

-  private final NuGetActionFactory 
- myNuGetActionFactory;
-  private final PackagesParametersFactory 
-  myParametersFactory;

-  public PackagesInstallerRunner(@NotNull 
-  final NuGetActionFactory nuGetActionFactory,
+ public class PackagesInstallerRunner extends 
+  NuGetRunnerBase {
+  public PackagesInstallerRunner(@NotNull 
+  final NuGetActionFactory actionFactory,
  @NotNull final PackagesParametersFactory 
  parametersFactory) {
-    myNuGetActionFactory = nuGetActionFactory;
-    myParametersFactory = parametersFactory;
+    super(actionFactory, parametersFactory);
  }

  @NotNull
@@ -72,7 +66,7 @@ private void createStages(@NotNull 
final BuildRunnerContext context,
            parameters,
            context.getBuild().getBuildLogger(),
            new PackagesInstallerBuilder(
-                    myNuGetActionFactory,
+                    myActionFactory,
                    stages,
                    context,
                    installParameters,
@@ -82,27 +76,8 @@ private void createStages(@NotNull 
final BuildRunnerContext context,
    stages.getLocateStage().pushBuildProcess(locate);
  }

-  @NotNull
-  public AgentBuildRunnerInfo getRunnerInfo() {
-    return this;
-  }

  @NotNull
  public String getType() {
    return PackagesConstants.INSTALL_RUN_TYPE;
  }

-  public boolean canRun(@NotNull 
-  BuildAgentConfiguration agentConfiguration) {
-    if (!agentConfiguration.getSystemInfo().
 -   isWindows()) {
      LOG.warn("NuGet packages installer available 
      only under windows");
      return false;
    }
\end{lstlisting}
\end{minipage}%
\hfill
%
\begin{minipage}[t]{.48\textwidth}
\begin{lstlisting}[caption={\textcolor{black}{PackagesPublishRunner Class}},label=PackagesPublishRunner,frame=single, breaklines=true,language=diff]{Name}@@ -85,8 +85,8 @@
- public class PackagesPublishRunner implements 
- AgentBuildRunner, AgentBuildRunnerInfo {
-  private static final Logger LOG =
  
-  Logger.getInstance(PackagesPublishRunner.
-  class.getName());

-  private final NuGetActionFactory myActionFactory;
-  private final PackagesParametersFactory 
-  myParametersFactory;

+ public class PackagesPublishRunner extends 
+ NuGetRunnerBase {
  public PackagesPublishRunner
  (@NotNull final NuGetActionFactory actionFactory,
  @NotNull final PackagesParametersFactory 
  parametersFactory) {
-    myActionFactory = actionFactory;
-    myParametersFactory = parametersFactory;
+    super(actionFactory, parametersFactory);
  }

  @NotNull
@@ -63,27 +58,8 @@ public void 
fileFound(@NotNull File file) throws RunBuildException {
    return process;
  }

-  @NotNull
-  public AgentBuildRunnerInfo getRunnerInfo() {
-    return this;
-  }

  @NotNull
  public String getType() {
    return PackagesConstants.PUBLISH_RUN_TYPE;
  }

-  public boolean canRun(@NotNull
-   BuildAgentConfiguration agentConfiguration) {
-    if (!agentConfiguration.getSystemInfo().
-     isWindows()) {
-      LOG.warn("NuGet packages installer available 
-      only under windows");
-     return false;
-   }
\end{lstlisting}
\end{minipage}%
\hfill
%

\end{figure*}

\begin{figure*}[htbp]

\begin{minipage}[t]{.48\textwidth}
\begin{lstlisting}[caption={\textcolor{black}{PackRunner Class}},label=PackRunner,frame=single,breaklines=true, language=diff]{Name}@@ -85,8 +85,8 @@
- public class PackRunner implements AgentBuildRunner, 
- AgentBuildRunnerInfo {
-  private static final Logger LOG = 
-  Logger.getInstance(PackRunner.class.getName());

-  private final NuGetActionFactory myActionFactory;
-  private final PackagesParametersFactory 
-  myParametersFactory;

+ public class PackRunner extends NuGetRunnerBase {
  public PackRunner(
    @NotNull final NuGetActionFactory actionFactory,
    @NotNull final PackagesParametersFactory 
    parametersFactory) {
-    myActionFactory = actionFactory;
-    myParametersFactory = parametersFactory;
+    super(actionFactory, parametersFactory);
  }

  @NotNull
@@ -54,28 +49,8 @@ public BuildProcess 
createBuildProcess(@NotNull final AgentRunningBuild 
runningB
    return process;
  }

-  @NotNull
-  public AgentBuildRunnerInfo getRunnerInfo() {
-    return this;
-  }

  @NotNull
  public String getType() {
    return PackagesConstants.PACK_RUN_TYPE;
  }

-  public boolean canRun(@NotNull 
-  BuildAgentConfiguration 
-    agentConfiguration) {
-    if (!agentConfiguration.getSystemInfo()
-    .isWindows()) {
-      LOG.warn("NuGet packages installer available 
-      only under windows");
-      return false;
-    }
-
-    if (!agentConfiguration.
-      getConfigurationParameters().
-    containsKey(DotNetConstants.
-    DOT_NET_FRAMEWORK_4_x86)) {
-      LOG.warn("NuGet requires 
-    .NET Framework 4.0 x86 installed");
-      return false;
-    }

-    return true;
-  }
\end{lstlisting}
\end{minipage}%
\hfill
%
\begin{minipage}[t]{.48\textwidth}
\begin{lstlisting}[caption={\textcolor{black}{NuGetRunnerBase Class}},label=NuGetRunnerBase,frame=single,breaklines=true, language=diff]{Name}@@ -85,8 +85,8 @@
/**
 * @author Eugene Petrenko (eugene.petrenko@gmail.com)
 *         Date: 23.08.11 18:32
 */
+ public abstract class NuGetRunnerBase implements 
+ AgentBuildRunner, AgentBuildRunnerInfo {
+  protected final Logger LOG = Logger.getInstance
+  (getClass().getName());

+  protected final NuGetActionFactory 
+ myActionFactory;
+  protected final PackagesParametersFactory
+  myParametersFactory;

+  public NuGetRunnerBase(NuGetActionFactory 
+ actionFactory, 
+  PackagesParametersFactory parametersFactory) {
+    myActionFactory = actionFactory;
+    myParametersFactory = parametersFactory;
+ }

+  @NotNull
+  public AgentBuildRunnerInfo getRunnerInfo() {
+    return this;
+  }

+  @NotNull
+  public abstract String getType();

+  public boolean canRun(@NotNull 
+  BuildAgentConfiguration 
+  agentConfiguration) {
+    if (!agentConfiguration.getSystemInfo()
+    .isWindows()) {
+      LOG.warn("NuGet packages installer available 
+      only under windows");
+      return false;
+    }

+    if (!agentConfiguration.
+    getConfigurationParameters().
+    containsKey(DotNetConstants.
+    DOT_NET_FRAMEWORK_4_x86)) {
+      LOG.warn("NuGet requires .NET Framework 
+      4.0 x86 installed");
+      return false;
+    }

+    return true;
+  }
+ }
\end{lstlisting}
\end{minipage}%

\end{figure*}


\subsubsection{\textcolor{black}{Visibility Changes.}}
\begin{figure*}[h]
\centering 
\includegraphics[width=1.0\textwidth]{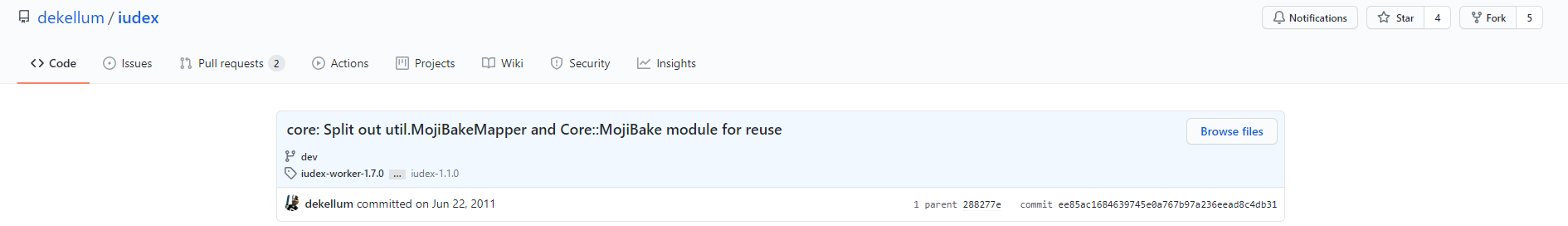}
\caption{
\textcolor{black}{Commit message stating the case of visibility changes \cite{dekellum}.}}
\label{fig:case study visibility}
\end{figure*}

\textcolor{black}{\textbf{\textit{Case study.}} The developers mention splitting up line of code:- util.MojiBakeMapper and Core::MojiBake for reuse in the commit message shown in Figure \ref{fig:case study visibility}. This refers to \texttt{mojibake.rb} file, Listing \ref{mojibake}, although the extract refactoring operation is seen to be performed on \texttt{MojiBakerFilter} and \texttt{MojiMapper} indicated by the Listing \ref{MojiBakeFilter}, respectively. The extract refactoring operation is performed on \texttt{MojiBakeFilter} from which the private method recover of type \texttt{CharSequence} was extracted to \texttt{MojiMapper} class as a public method recover of type \texttt{CharSequence}. Although the commit message suggests reuse, the refactoring operation is performed on another file, which can be considered as a case study for visibility changes as the access modifier of the method recover of type \texttt{Charsequence} was changed from private to public. On further investigation, we searched for any traces of reusability of the method recover being extracted and found a few classes that used recover method. 
}

\begin{figure*}[htbp]
\noindent\begin{minipage}[t]{.48\textwidth}
\begin{lstlisting}[caption={\textcolor{black}{Before refactoring}},label=BFBEFORE, frame=single,breaklines=true, language=diff]{Name}@@ -85,8 +85,8 @@
-  if (c == 27
-    && pushBackChar.isEmpty()
-    && in.isNonBlockingEnabled()
-    && in.peek(escapeTimeout) == -2) {
-        Object otherKey = ((KeyMap) o).
-         getAnotherKey();
-  if (otherKey == null) {
-        otherKey = ((KeyMap) o).
-        getBound(Character.toString((char) c));}
-        o = otherKey;
-  if (o == null || o instanceof KeyMap) {
-                            continue; }
-        sb.setLength(0);  }
-  else {                    continue; }
 *
 *
-   while ( o == null && sb.length() > 0 ) {
-    c = sb.charAt( sb.length() - 1 );
-        sb.setLength( sb.length() - 1 );
-        Object o2 = getKeys().getBound( sb );
-        if ( o2 instanceof KeyMap ) {
-            o = ((KeyMap) o2).getAnotherKey();
-        if ( o == null ) {
-                     continue;
-        } else {
-                     pushBackChar.push( (char) c ); 
-        }}
*
*
*
*
*
*
\end{lstlisting}
\end{minipage}%
\hfill
%
\begin{minipage}[t]{.48\textwidth}
\begin{lstlisting}[caption={\textcolor{black}{After refactoring}},label=BFAFTER,frame=single, breaklines=true, language=diff]{Name}@@ -85,8 +85,8 @@
+  if (c == ESCAPE
+    && pushBackChar.isEmpty()
+    && in.isNonBlockingEnabled()
+    && in.peek(escapeTimeout) == READ_EXPIRED) {
+        Object otherKey = ((KeyMap) o).
+        getAnotherKey();
+    if (otherKey == null) {
                        
+          otherKey = ((KeyMap) o).getBound(Character.
+                    toString((char) c));
+                    }
+                    o = otherKey;
+                    if (o == null || 
+                    o instanceof KeyMap) {
+                        continue;
+                    }
+                    opBuffer.setLength(0);
+                } else {
+                    continue;
+                }
*
*
+    while (o == null && opBuffer.length() > 0) {
+      c = opBuffer.charAt(opBuffer.length() - 1);
+           opBuffer.setLength(opBuffer.length() - 1);
+                Object o2 = keys.getBound(opBuffer);
+                if (o2 instanceof KeyMap) {
+                    o = ((KeyMap) o2).getAnotherKey();
+                    if (o == null) {
+                        continue;
+                    } else {
+                        pushBackChar.push((char) c);
+                    }}

\end{lstlisting}
\end{minipage}%

\end{figure*}

\begin{figure*}[htbp]
\noindent\begin{minipage}[t]{.48\textwidth}
\begin{lstlisting}[caption={\textcolor{black}{PubSubElementProcessorAbstract Class}},label=PubSubElementProcessorAbstract, frame=single,breaklines=true,language=diff]{Name}@@ -85,8 +85,8 @@

+ public abstract class PubSubElementProcessorAbstract
+    implements PubSubElementProcessor
+ {

+    protected BlockingQueue<Packet> outQueue;
+    protected DataStore             dataStore;
+    protected Element               element;
+    protected IQ                    response;
+    protected IQ                    request;
+    protected JID                   actor;
+    protected String                serverDomain;
+    protected String                topicsDomain;
+    protected String                node;
+    protected Helper    configurationHelper;
+	public void setOutQueue(BlockingQueue<Packet> 
+	outQueue)
+	{
+		this.outQueue = outQueue;
+	}
+	public void setDataStore(DataStore dataStore)
+	{
+		this.dataStore = dataStore;
+	}
+	public void setServerDomain(String domain)
+	{
+		serverDomain = domain;
+	}
*
*
*
*
*
*
*
*
\end{lstlisting}
\end{minipage}%
\hfill
%
\begin{minipage}[t]{.48\textwidth}
\begin{lstlisting}[caption={\textcolor{black}{NodeCreate Class}},label=NodeCreate,frame=single,breaklines=true, language=diff]{Name}@@ -85,8 +85,8 @@
- public class NodeCreate implements 
- PubSubElementProcessor
+ public class NodeCreate extends 
+ PubSubElementProcessorAbstract
{
-    private static final Logger LOGGER = 
-    Logger.getLogger(NodeCreate.class);
-    private BlockingQueue<Packet> outQueue;
-    private DataStore             dataStore;
-	private Element               element;
-    private IQ                    response;
-    private IQ                    request;
-    private JID                   actor;
-	private String                serverDomain;
-	private String                topicsDomain;
-	private String                node;
-	private Helper     configurationHelper;
	private static final Pattern nodeExtract = Pattern
	.compile("^/user/[^@]+@([^/]+)/[^/]+$");
    private static final String NODE_REG_EX  
    = "^/user/[^@]+@[^/]+/[^/]+$";
@@ -46,16 +36,6 @@ public NodeCreate
(BlockingQueue<Packet> outQueue, DataStore dataStore)
    	setOutQueue(outQueue);
    }

-	public void setOutQueue(BlockingQueue<Packet> 
-	outQueue)
-	{
-		this.outQueue = outQueue;
-	}
-
-	public void setDataStore(DataStore dataStore)
-	{
-		this.dataStore = dataStore;
-	}

\end{lstlisting}
\end{minipage}%

\end{figure*}

\begin{tcolorbox}
\textbf{Summary.} Code reuse helps wit various software development activities. Our analysis found that the extraction of reusable component leads the rationale behind reuse-related code changes. Refactoring reusable code also helps with API management, duplicate code removal, changes to visibility, test code reorganization, and implementing design patterns. 
\end{tcolorbox}

\section{Implications}\label{sec:implication}
The main implications of this study are as follows:
\begin{itemize}
    \item 
   
    \textcolor{black}{\textbf{Further exploiting quality metrics and reusability refactoring.} The existing literature discusses different automatic refactoring approaches that help practitioners in detecting anti-patterns or code smells. More recently, Baqais and Alshayeb \cite{baqais2020automatic} show that there is an increase in the number of studies on automatic refactoring approaches and researchers have begun exploring how machine learning can be used in identifying refactoring opportunities. Since the features play a vital role in the quality of the obtained machine learning models, this study can help determine which metrics can be used as effective features in machine learning algorithms to accurately predict refactoring opportunities at different levels of granularity (i.e., class, method, field), which can assist developers in automatically making their decisions. For example, using the most impactful metrics as a feature to predict whether a given piece of code should undergo a specific refactoring operation makes developers more confident in accepting the recommended refactoring or picking out the most suitable reusable candidate. Such knowledge is needed as, in practice, the built model should require as little data as possible. Further, since we observe from RQ2 that most of the reusability metrics did not capture any improvement, we plan to conduct more experiments to validate the effectiveness of reusability metrics to explore if the observations are due to the appropriateness of the reusability quality metrics or to the needed validation and clarity of developers perception.}
    
    \textbf{Reducing the amount of efforts to refactor the code to improve its  reusability.} Generally, reused classes tend to be more maintainable than native classes. One particular aspect of refactoring is to increase the reusability of software components. However, a recent study \cite{feitosa2020code} found that the reused code is in need for various refactorings even though the produced code obeys to good object-oriented practices. Our study sheds light on developers' strategies to refactor the code to improve its reusability that is different from refactoring applied in mainstream development (e.g., reusability refactorings heavily impact methods while typical refactorings, impact all code elements). Understanding such strategies assist in providing developers with a more efficient way to utilize existing code to create new functionality, and facilitate development and maintenance since less work is needed to accomplish additional functionality.
    \begin{figure}[htbp]
\noindent\begin{minipage}[t]{.48\textwidth}
\begin{lstlisting}[caption={\textcolor{black}{Before refactoring - In SerializingTranscoderTest.java)}},label=Test1 (before), frame=single,breaklines=true,language=diff]{Name}@@ -85,8 +85,8 @@
-   public void testLong() throws Exception {
-		assertEquals(923l, tc.decode(tc.encode(923l)));
-	}
-	public void testInt() throws Exception {
-		assertEquals(923, tc.decode(tc.encode(923)));
-	}
-	public void testChar() throws Exception {
-		assertEquals('c', tc.decode(tc.encode('c')));
-	}
-	public void testBoolean() throws Exception {
-		assertSame(Boolean.TRUE, 
-       tc.decode(tc.encode(true)));
-		assertSame(Boolean.FALSE, 
-       tc.decode(tc.encode(false)));
-	}
\end{lstlisting}
\end{minipage}%
\hfill
%
\begin{minipage}[t]{.48\textwidth}
\begin{lstlisting}[caption={\textcolor{black}{Before refactoring - In WhalinTranscoderTest.java}},label=Test2 (before),frame=single,breaklines=true, language=diff]{Name}@@ -85,8 +85,8 @@
-   public void testLong() throws Exception {
-		assertEquals(923l, tc.decode(tc.encode(923l)));
-	}
-	public void testInt() throws Exception {
-		assertEquals(923, tc.decode(tc.encode(923)));
-	}
-	public void testShort() throws Exception {
-		assertEquals((short)923, tc.decode(
-       tc.encode((short)923)));
-	}
-	public void testChar() throws Exception {
-		assertEquals('c', tc.decode(tc.encode('c')));
-	}
-	public void testBoolean() throws Exception {
-		assertSame(Boolean.TRUE, 
-       tc.decode(tc.encode(true)));
-		assertSame(Boolean.FALSE, 
-       tc.decode(tc.encode(false)));
-	}

\end{lstlisting}

\end{minipage}%

\hfill
%
\begin{minipage}[t]{.48\textwidth}
\begin{lstlisting}[caption={\textcolor{black}{After refactoring - In BaseTranscoderCase.java}},label=Test (after),frame=single, breaklines=true,language=diff]{Name}@@ -85,8 +85,8 @@
+   public void testLong() throws Exception {
+		assertEquals(923l, tc.decode(tc.encode(923l)));
+	}
+	public void testInt() throws Exception {
+		assertEquals(923, tc.decode(tc.encode(923)));
+	}
+	public void testShort() throws Exception {
+		assertEquals((short)923, 
+       tc.decode(tc.encode((short)923)));
+	}
+	public void testChar() throws Exception {
+		assertEquals('c', tc.decode(tc.encode('c')));
+	}
+	public void testBoolean() throws Exception {
+		assertSame(Boolean.TRUE, 
+       tc.decode(tc.encode(true)));
+		assertSame(Boolean.FALSE, 
+       tc.decode(tc.encode(false)));
+	}

\end{lstlisting}

\end{minipage}%

\end{figure}

    \item 
    \textbf{Examining the code reuse potentials with refactoring}. Our study reveals the context in which developers refactor the code for the purpose of improving code reusability. Our future research direction can focus on providing a comprehensive taxonomy of reusability-aware refactorings. This taxonomy can show various contexts of reusability refactoring and demonstrate different forms of reuse. Thereafter, researchers can build on top of our RQ3 findings to better understand developers practices and investigate to what extent this reusability-aware refactoring taxonomy improves the quality of the system.
    \item 
    \textbf{Understanding the completeness of the quality metrics in capturing the reusability improvements as documented by developers.} We observed that not all of the quality metrics are able to capture the reusability improvement as perceived by developers in their commit messages. While quality metrics can help pinpoint design flaws for refactoring recommendation systems, such recommendation would be meaningful if it is complemented with qualitative insights from developers.
    \item 
    \textbf{Extending/varying the basic structure of design patterns to support real-world applicability.} Our qualitative analysis (see Figure \ref{fig:case study design pattern commit message}) shows that developers in practice are not following the exact basic structure of the design patterns that usually appears in the original documentation of the well-known catalog of patterns \cite{gamma1995elements}. Developers instead alter the implemented design patterns according to the developer's needs. Future researchers are encouraged to perform a deeper investigation on understanding the mismatch between theory and practice, and propose an approach that can detect not only patterns in their basic form but also modified versions of patterns. 
   \end{itemize}

\section{Threats to Validity}\label{sec:threat}

\noindent\textcolor{black}{\textbf{Internal Validity.}
We analyzed only the 28 refactoring operations detected by Refactoring Miner, which can be viewed as a validity threat because the tool did not consider all refactoring types mentioned by Fowler et al. \cite{fowler2018refactoring}. However, in a previous study, Murphy-Hill et al. \cite{murphy2012we} reported that these types are amongst the most common refactoring types. Moreover, we did not perform a manual validation of refactoring types detected by Refactoring Miner to assess its accuracy, so our study is mainly threatened by the accuracy of the detection tool. Yet, Tsantalis et al. \cite{tsantalis2018accurate} reported that Refactoring Miner has a precision of 98\% and a recall of 87\% which significantly outperforms the previous state-of-the-art tools, which gives us confidence in using the tool. Another threat to validity is that, as we mentioned above, while we determined whether a commit has a reusability change, we only look for terms like \textit{reus} in the commit message, although not all reusability commit messages may contain those words. Another critical threat, is the fact that not all refactorings are root-canal. Developers may be interleaving refactorings with other types of changes, and so, this may become a noise in our measurements. To mitigate this issue, we considered commits that both contain an explicit statement about reusability, and contain at least one refactoring operation, in order to correlate between the refactoring and its documentation. Also, the existence of several unrelated files, in the commit, as part of other changes, can also become a noise for our metrics measurements. To mitigate this threat, we measure the metrics for code elements that are being refactored, and not all the changed files in the reusability commit.}

\noindent\textcolor{black}{\textbf{External Validity.} The first threat is that the analysis was restricted to only open source, Java-based,
Git-based repositories. However, we were still able to analyze 1,828 projects that are highly varied
in size, contributors, number of commits and refactorings. }

\noindent\textcolor{black}{\textbf{\textcolor{black}{Construct Validity.}} A potential threat to construct validity
relates to the set of metrics, as it may miss some properties of
the selected internal quality attributes. To mitigate this threat,
we select well-known metrics that cover various properties of
each attribute, as reported in the literature \cite{chidamber1994metrics}.}

\textcolor{black}{While our experiments rely on mining the intention of developers through their explicit documentation in code, which is in line with what has been done by various recent studies \cite{pantiuchina2018improving,alomar2019impact,alomar2020exploratory,paixao2020behind,fernandes2020refactoring,hamdi2021empirical}, this may not cover the whole spectrum of all the code changes done with reuse in mind. Thus, we might be missing some code changes that were performed with that aspect but without any explicit documentation about it (i.e., false negatives). Therefore, it would be interesting to further investigate the impact of reusability on code changes by interviewing developers about it.}

 \begin{figure}[htbp]
\noindent\begin{minipage}[t]{.48\textwidth}
\begin{lstlisting}[caption={\textcolor{black}{mojibake.rb}},label=mojibake, frame=single,breaklines=true,language=diff]{Name}@@ -85,8 +85,8 @@
require 'iudex-core'
require 'java'

-module Iudex::Core::Filters
-  import 'iudex.core.filters.MojiBakeFilter'
-
-  # Re-open iudex.core.filters.MojiBakeFilter to add 
-  config file
-  # based initialization.
-  class MojiBakeFilter
+ module Iudex::Core

+  module MojiBake
    DEFAULT_CONFIG = File.join( File.dirname
    ( __FILE__ ),
                                '..', '..', 'config',
                                'mojibake' )
\end{lstlisting}
\end{minipage}%
\hfill
%
\begin{minipage}[t]{.48\textwidth}
\begin{lstlisting}[caption={\textcolor{black}{MojiBakeFilter Class}},label=MojiBakeFilter,frame=single,breaklines=true, language=diff]{Name}@@ -85,8 +85,8 @@
-    private CharSequence recover( CharSequence in )
-    {
-        Matcher m = _mojiPattern.matcher( in );
-        StringBuilder out = new StringBuilder( 
-        in.length() );
-        int last = 0;
-       while( m.find() ) {
-            out.append( in, last, m.start() );
-            String moji = in.subSequence( m.start(), 
-            m.end() ).toString();
-            out.append( _mojis.get( moji ) );
-            last = m.end();
-        }
-        out.append( in, last, in.length() );

-        if( out.length() < in.length() ) {
-           return recover( out );
-        }
-        else {
-            return out;
-        }
-    }

-    private final Key<CharSequence> _field;
-
-    private final Pattern _mojiPattern;
-    private final HashMap<String, String> _mojis;
+    private final MojiBakeMapper _mapper;
}

\end{lstlisting}

\end{minipage}%

\hfill
%
\begin{minipage}[t]{.48\textwidth}
\begin{lstlisting}[caption={\textcolor{black}{MojiBakeMapper Class}},label=MojiBakeMapper,frame=single,breaklines=true, language=diff]{Name}@@ -85,8 +85,8 @@
+ public class MojiBakeMapper
+ {
+    public MojiBakeMapper( String regex,
+                           Map<String,String> mojis )
+    {
+        _mojiPattern = Pattern.compile( regex );
+        _mojis = new HashMap<String,String>( mojis );
+    }
+    public CharSequence recover( CharSequence in )
+    {
+        Matcher m = _mojiPattern.matcher( in );
+        StringBuilder out = new StringBuilder( 
+        in.length() );
+        int last = 0;
+        while( m.find() ) {
+            out.append( in, last, m.start() );
+            String moji = in.subSequence( m.start(),
+            m.end() ).toString();
+            out.append( _mojis.get( moji ) );
+            last = m.end();
+        }
+        out.append( in, last, in.length() );

+        if( out.length() < in.length() ) {
+            return recover( out );
+        }
+        else {
+            return out;
+        }
+   }
+    private final Pattern _mojiPattern;
+    private final HashMap<String, String> _mojis;
+ }

\end{lstlisting}

\end{minipage}%

\end{figure}

\section{Conclusion} \label{sec:conclusion}
In this paper, we performed a study on analyzing reusability refactorings based on information in Java projects from our dataset. We found that in reusability refactorings, the changes developers performed would significantly affect metrics pertaining to methods, but not significantly affect metrics regarding comments or cohesion of classes. We also found that less than 0.4\% commits are reusability refactorings in 154,820 commits. Another fact we found is that 
method is modified more frequently in reusability refactoring changes. Our results have shown some existing facts in reusability refactorings, and those findings could help developers to make better decisions while performing reusability refactorings in the future.

Some recommendations that we have for future work involve comparing different subsections of data, 
 and determining what refactorings are related to reusability. Specifically, we think that it would be interesting to compare the results that we got to instances where each individual refactoring detected was analyzed to explore if it was done for reusability or not, to see if us grouping all refactorings in a commit for reusability and non-reusability is similar. We also think that analyzing the code before and after the reusability commits for different metrics that are more usability based, such as adaptability, understandability, or portability, could be an interesting future work, though an issue might arise to finding specific ways to measure those metrics. 
 Moreover, we plan to find a better way to figure out if a commit was a reusability refactoring or not. Since this work relies on the commit message, there could be commits incorrectly labeled, or commits that are reusability but not labeled as such that we are missing.
 
 \textcolor{black}{Further, we designed our study with the goal of better understanding developer perception of code reusability within the open source community. Further research in this regard is needed (e.g.,  running contextual interviews with developers to uncover the underpinning reasons for code reuse during refactorings). As with every study, the results may not generalize to other contexts. Extending this work with the industry partners is part of our future investigation to challenge our current findings.}

\bibliographystyle{spmpsci}
\bibliography{MyBibliography.bib}

\end{document}